\documentclass[conference]{IEEEtran}
\IEEEoverridecommandlockouts
\usepackage{cite}
\usepackage{array}
\usepackage{amsmath,amssymb,amsfonts}
\usepackage{algorithmic}
\usepackage{graphicx}
\usepackage{textcomp}
\usepackage{xcolor}
\usepackage{booktabs}
\usepackage{subcaption}
\usepackage{multirow}
\usepackage[most]{tcolorbox}
\usepackage[dvipsnames]{xcolor}
\usepackage{enumitem}
\usepackage{colortbl}
\usepackage{tikz}
\usepackage{url}
\usepackage{hyperref}
\usepackage{dblfloatfix}
\usepackage{pifont}
\setlist{noitemsep, topsep=2pt, parsep=0pt, partopsep=0pt}
\makeatletter
\newcommand{\linebreakand}{%
  \end{@IEEEauthorhalign}
  \hfill\mbox{}\par
  \mbox{}\hfill\begin{@IEEEauthorhalign}
}
\makeatother

\def\BibTeX{{\rm B\kern-.05em{\sc i\kern-.025em b}\kern-.08em
    T\kern-.1667em\lower.7ex\hbox{E}\kern-.125emX}}

\renewcommand{\arraystretch}{0.9}

\newcommand{\Vacation}{%
{\textbf{Vacation }}%
}

\newcommand{\Movie}{%
{\textbf{Movie }}%
}

\newcommand{\Analysis}{%
\tcbox[
    enhanced,
    on line,
    box align=base,
    nobeforeafter,
    colback=Apricot!60,
    colframe=Apricot!60,
     left=1pt,
    right=1pt,
    top=1pt,
    bottom=1pt,
    boxsep=0.4pt
]{\textbf{Analysis}}%
}

\newcommand{\Creation}{%
\tcbox[
    enhanced,
    on line,
    box align=base,
    nobeforeafter,
    colback=OliveGreen!35,
    colframe=OliveGreen!35,
     left=1pt,
    right=1pt,
    top=1pt,
    bottom=1pt,
    boxsep=0.4pt
]{\textbf{Creation}}%
}

\newcommand{\Budget}{%
{\textbf{Budget }}%
}

\newcommand{\Schedule}{%
{\textbf{Schedule }}%
}

\begin{document}

\title{Plans Work in Mysterious Ways: \\Evaluating a Plan Mode for Spreadsheet Agents}

\author{
\IEEEauthorblockN{Aayush Kumar\IEEEauthorrefmark{1}, Avik Dutta\IEEEauthorrefmark{1}, Sumit Gulwani\IEEEauthorrefmark{2}, Gustavo Soares\IEEEauthorrefmark{2}, Advait Sarkar\IEEEauthorrefmark{3}, Emerson Murphy-Hill\IEEEauthorrefmark{4}}
\IEEEauthorblockA{
\textit{Microsoft}\\
\IEEEauthorrefmark{1}Bengaluru, India \quad
\IEEEauthorrefmark{2}Redmond, WA, USA \quad
\IEEEauthorrefmark{3}Cambridge, UK \quad
\IEEEauthorrefmark{4}Sunnyvale, CA, USA\\
\{t-aaykumar, avikdutta, sumitg, gustavo.soares, advait, emerson.rex\}@microsoft.com}
 }

\maketitle

\begin{abstract}
Plan Modes have become standard features in agentic programming tools, allowing users to gain transparency and control by working with the agent to develop a plan before task execution. However, it remains unclear whether the benefits of this feature translate to end-user programming environments such as spreadsheets. Since spreadsheet programmers tend to work iteratively and care less about technical correctness, upfront planning may not fit into their workflows as easily. In this paper, we build a prototype of a Plan Mode for spreadsheet programming and evaluate it against a non-planning baseline through a within-subjects user study (N=24). We found that despite similar task outcomes with both tools, using Plan Mode led to a reduction in refinement and a better perception of the tool across dimensions of creativity support and human-machine collaboration. We discuss the implications of these results for the future design of Plan Modes, and for the broader role of human-AI planning in end-user programming.

\end{abstract}


\begin{figure*}[!t]
\centering
\vspace{-2em}
\begin{subfigure}[t]{0.31\textwidth}
    \centering
    \includegraphics[width=\linewidth]{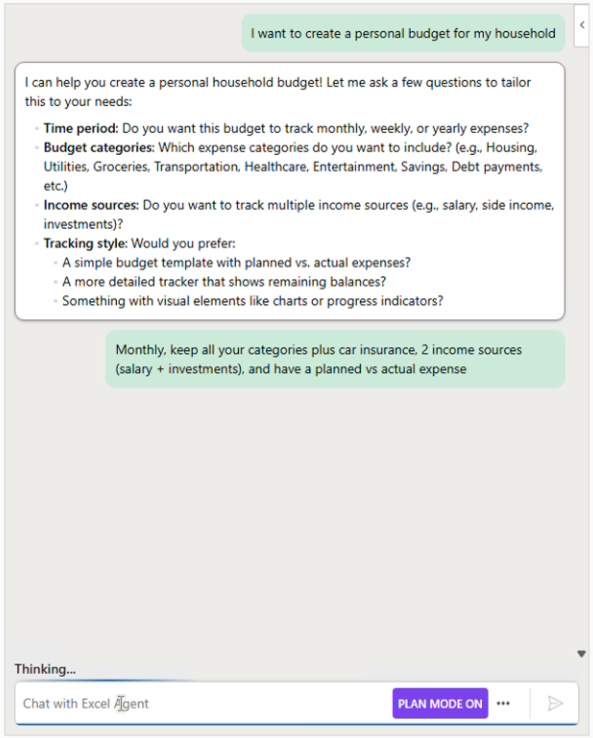}
    \caption{The Plan agent asks clarifying questions to resolve any ambiguities in the user's query}
    \label{fig:sub1}
\end{subfigure}
\hfill
\begin{subfigure}[t]{0.31\textwidth}
    \centering
    \includegraphics[width=\linewidth]{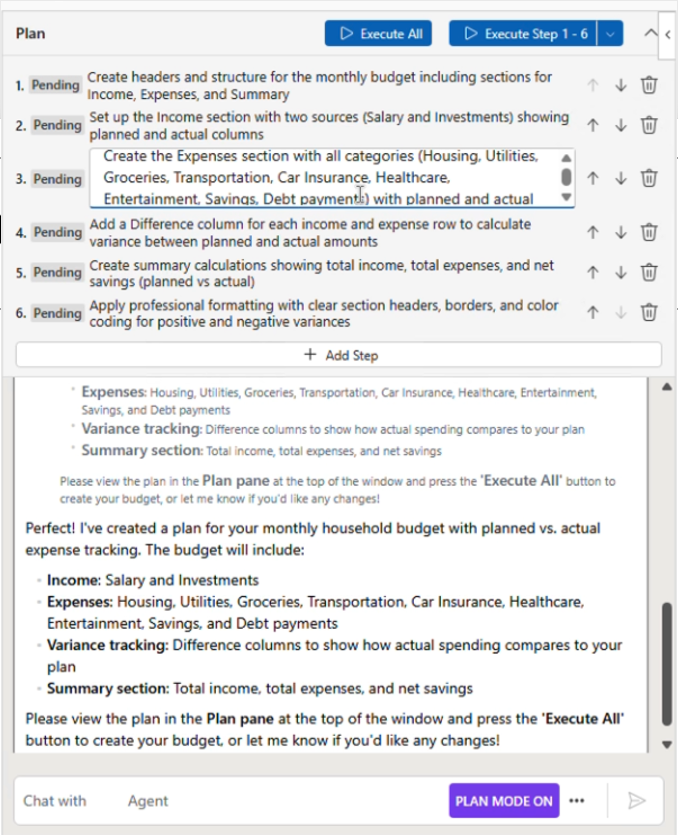}
    \caption{Users can directly manipulate the plan artifact}
    \label{fig:sub2}
\end{subfigure}
\hfill
\begin{minipage}[t]{0.31\textwidth}
    \centering
    \vspace{-195pt}

    \begin{subfigure}[t]{\linewidth}
        \centering
        \includegraphics[width=\linewidth]{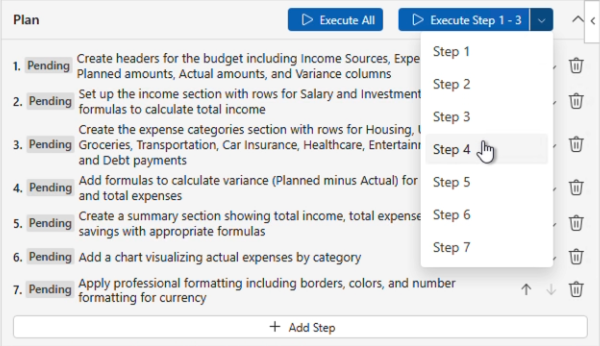}
        \caption{Partial execution allows users to evaluate the plan at intermediate stages}
        \label{fig:sub3}
    \end{subfigure}

    \vspace{0.2em}

    \begin{subfigure}[t]{\linewidth}
        \centering
        \includegraphics[width=\linewidth]{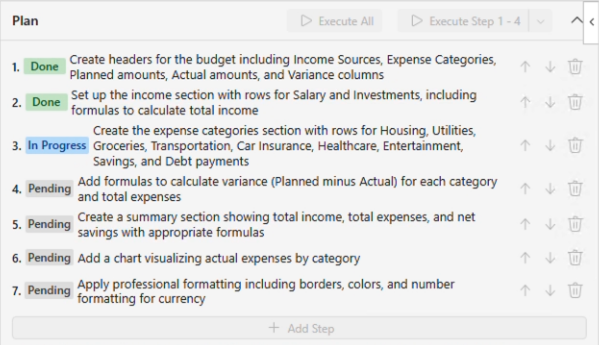}
        \caption{The Act agent updates the status of the plan during execution}
        \label{fig:sub4}
    \end{subfigure}
\end{minipage}
\vspace{-1.5em}
\caption{Interacting With Plan Mode}
\vspace{-1em}
\label{fig:combined}
\end{figure*}

\section{Introduction}

AI tools are getting dramatically better at performing complex tasks autonomously. \textit{Agentic} tools -- tools that interact with their environment and iterate on their own outputs -- exemplify this progress. But as these capabilities grow, so do the challenges users face in specifying task goals and communicating preferences to such tools \cite{hac_challenges}. To address these challenges, agentic coding tools such as Claude Code\cite{claude_code}, VSCode Agent Mode\cite{vscode_plan_agent}, Cursor\cite{cursor_plan_mode}, and Cline\cite{cline_plan_act} have adopted a \textit{Plan Mode}: programmers work with the agent to develop a plan before it begins executing the task. This lets programmers control the agent's implementation approach and see the edits the agent will make before it makes them.

While Plan Mode has successfully become a part of the programmer's toolkit, it remains unclear whether this affordance ports over smoothly to end-user programming domains such as spreadsheet programming. 
Unlike professional programmers, end-user programmers tend to work more iteratively, redefining their task constraints in an emergent manner as they work through their tasks \cite{euse}. End-user programmers may also lack the computational thinking skills needed to 
work with AI tools effectively \cite{sarkar2022likeprogramartificialintelligence}. Further, while one of the primary benefits of Plan Mode for programmers is gaining transparency into the technical implementation approach taken by the agent, spreadsheet programmers care less about technical correctness and more about the practical utility of the workbook \cite{half_measures}.

On the other hand, an interactive planning mode may also have benefits specific to spreadsheet programmers. Since end-user programmers struggle to turn ideas into working programs \cite{euse}, planning can help scaffold complex spreadsheet tasks. Prior work indicates that planning-related affordances such as shared artifacts editable by both agent and user \cite{help_me_think, steering_and_verification} and clarifying questions asked by the agent \cite{dango, promptions, tabletalk} can improve user experiences when using AI tools in spreadsheet programming and related domains such as open-ended data analysis. Despite this, the impact of planning affordances on spreadsheet task outputs remains unclear from prior work. Further, to our knowledge, explicit planning modes for agentic spreadsheet programming tools remain unexplored in prior research. 
In this paper, we begin to address these gaps by 1) articulating a set of design features for plan modes in spreadsheet programming and implementing these features in a prototype, and 2) evaluating the effects of our Plan Mode prototype on process, outcomes, and experience in complex open-ended spreadsheet tasks. 

In a controlled mixed-methods study (N=24) comparing Plan Mode with a baseline agentic tool (`Act Mode') for spreadsheet programming, we found that Plan Mode changed how participants exchanged information with the tool, shifting requirements elicitation from iterative refinement toward responding to clarifying questions. We also found that participants preferred Plan Mode over the baseline across dimensions of creativity support and human-machine collaboration. Despite this, we did not find any notable differences in the final workbooks created by participants, indicating that the benefits of Plan Mode primarily emerge from its interaction mechanisms, such as clarifying questions and UI elements.

\section{Related Work}
\subsection{Interactive Human-AI Planning}
Prior work has explored interactive tools for human-AI collaborations on technically complex tasks across domains such as programming \cite{ bridging_abstraction_gap, pail, why_ai_agents, biscuit}, data analysis \cite{dataspeck, dango, promptions, 10.1145/3663384.3663389, steering_and_verification}, and financial tasks \cite{help_me_think, plan_then_execute}. 
Many of these tools explicitly or implicitly involve some form of interactive planning, particularly when the AI tool is an \textit{agent} (that is, it is capable of making complex programs autonomously). Prior research suggests that planning can help users communicate their expectations of how a task should be performed and gain transparency into the agent's actions \cite{magentic_ui}.
Such planning affordances are often reified through shared artifacts editable by users and AI agents that represent actions taken/to be taken by the agent. For example, He et al. \cite{plan_then_execute} and Mozannar et al. \cite{magentic_ui} introduce tools with an iterative loop where the agent proposes a plan that users can update before execution. 
Cocoa\cite{cocoa} maintains a similar shared plan, but further allows for iteration on the plan at intermediate stages during execution.  

Planning affordances have become prominent for coding tools in particular. While planning collaboratively with coding agents enables developers to mitigate risk and ensure that the agent's implementation approach aligns with their expectations \cite{dong2025correctnesscollaborationhumancenteredframework}, even non-collaborative planning by the LLM itself can improve performance on coding tasks \cite{codeplan}. This has led to the adoption of Plan Modes in many popular coding agents, such as Claude Code\cite{claude_code}, VSCode Agent Mode\cite{vscode_plan_agent} and Cursor\cite{cursor_plan_mode}.

Despite this, prior work is mixed on the benefits of explicit planning stages in human-AI collaborations, indicating that while planning can lead to higher task clarity, transparency, and steerability, it can also add cognitive burdens for users \cite{steering_and_verification}. Further, users can mistrust incorrect plans that look plausible \cite{plan_then_execute}. Such errors can arise from a lack of tacit knowledge and a failure to account for user assumptions \cite{passi2025agentic,cocoa,why_ai_agents}.
In this paper, we extend this line of work to end-user spreadsheet programming by comparing a Plan Mode for spreadsheet agents to a non-planning baseline.

Clarifying questions asked by agents to gain information about the user's task have emerged as an important affordance for improved experiences. Such questions represent a more implicit form of planning where the agent identifies and resolves ambiguities in a task before execution, as opposed to proceeding with execution based on assumptions and expecting the user to flag issues post-hoc.
For example, in PAIL \cite{pail}, users must answer questions related to design choices before proceeding with implementation. The authors of Dango\cite{dango} and DataSpeck\cite{dataspeck} report that asking clarifying questions led to improved intent clarification and efficiency in data tasks. Drosos et al. \cite{promptions} report that dynamic prompt refinement through UI-based clarifications improves user experiences. Similarly, Cheng et al. \cite{biscuit} found that UI-based interactions during task execution make it easier for participants to explore new ideas in open-ended coding tasks. Yet, it remains unclear whether and how such clarifying questions affect the quality of task outputs\cite{pail, promptions}. In this work, we begin to fill this gap by measuring participant outputs across a clarifying question-based Plan Mode and a baseline.

\subsection{AI tools for Spreadsheet Programming}
Prior work reveals a variety of AI-based spreadsheet programming tools with different interaction affordances. SheetCopilot\cite{sheetcopilot} and SheetAgent\cite{sheetagent} are autonomous AI tools capable of performing complex spreadsheet tasks. The Invisible Mentor\cite{litao} 
provides spreadsheet users with suggestions for their workflow based on screen recordings, while the Table Illustrator \cite{illustrator} interacts with participants using a puzzle-based metaphor to build plain tables. Other AI tools use some form of interactive intent clarification -- Drosos et al. \cite{promptions} use a UI for prompt refinement, TableTalk \cite{tabletalk} gathers feedback from users at regular stages of spreadsheet development, and Liu et al. \cite{bridging_abstraction_gap} use editable task-decomposition based plan representations for limited-scope tasks. However, to the best of our knowledge, as of June 2026, no existing spreadsheet agent contains an explicit Plan Mode that asks clarifying questions to create a persistent plan artifact (we exclude Excel Copilot \cite{excel_copilot} here since its Plan Mode is based on the findings of this work). For example, while Shortcut AI \cite{shortcut_ai} contains an explicit planning mode, this feature aims to translate user intent into a detailed implementation-heavy prompt rather than a persistent plan, and while Claude for Excel \cite{claude_for_excel} can ask clarifying questions and present a `plan' within its response for execution, this plan is not persistent and this feature is triggered only implicitly and at the agent's discretion. In this paper, we begin to fill this gap by building and evaluating one such Plan Mode for a spreadsheet programming agent.

Despite the prominence of Plan Modes for programming, there is reason to believe human-AI planning may have different results when done by end-user programmers rather than by professional programmers. 
Ko et al. \cite{euse} report that the benefits of explicit specifications might be unclear with end-user programmers. Since end-user programmers' primary motivations are their own personal goals, not the software itself, they tend to discover their requirements in an emergent fashion during execution. Similarly, Pandita et al. \cite{half_measures} report that spreadsheet programmers tend to build spreadsheets in a piecemeal manner, prioritizing practical use over technical correctness. Thus, upfront planning may not align with users' workflows, especially for complex tasks. On the other hand, interactive planning may also have its own unique benefits for end-user programmers. 
Sarkar et al. \cite{sarkar2022likeprogramartificialintelligence} report that using AI coding tools effectively requires computational thinking skills that end-user programmers might not have, such as decomposing problems into subproblems. Plan Mode can support users in building and enacting such skills. Similarly, Plan Mode can act as an affordance to translate users' ideas into workable programs, something that end-users struggle with \cite{euse}. Since end-user programmers find it difficult to comprehend AI outputs and spreadsheets not authored by themselves \cite{comprehension,will_code_remain}, Plan Mode may also make it easier for end-users to verify the agent's output. 
In this paper, we test these hypotheses by measuring participants' experiences and outputs when using Plan Mode for complex open-ended spreadsheet tasks.

\section{The Plan Mode Prototype}
In an agentic tool that can perform technically complex tasks autonomously, `Plan Mode' refers to a feature that allows users to
retain control and transparency over the agent's actions by developing a plan before task execution. This usually involves an interactive planning process which culminates in the agent presenting a plan for users to review based on the user's query, context, and responses to clarifying questions asked by the agent, allowing users to catch any incorrect assumptions early.
After agreeing on a plan, the agent leaves Plan Mode and starts executing the task adhering to this plan, providing users with transparency into the edits the agent makes before it makes them. 
In this section, we describe how our prototype for Plan Mode for spreadsheet development in Microsoft Excel implements these goals through an illustrative example (\S \ref{example}), the instructions we provided to the agent (\S \ref{instructions}), the design features of our tool (\S \ref{design}), and the tool's underlying implementation (\S \ref{implementation}).

\subsection{Illustrative Example} \label{example}
Khushi is creating a spreadsheet to keep track of her household expenses. She decides to use Plan Mode for this task, and submits the query \textit{``I want to create a personal budget for my household"}.
The agent replies with a list of clarifying questions about her budget, such as the time period and income sources. Based on her personal context and taste, Khushi replies to all these questions, skipping a sub-question on visualizations that she doesn't have an opinion on (Figure \ref{fig:sub1}).
Based on these answers, the agent builds and presents a plan to build the budget. On reviewing the plan, Khushi notices that the Expenses section includes an irrelevant category for debt payments, and so decides to remove this category by editing the corresponding step in the plan (Figure \ref{fig:sub2}). She further decides that she wants a chart to track spending across different categories, and expresses this in the chat to the agent, which then updates the plan to reflect this. Now satisfied with the plan, Khushi decides to execute it up to step 4 (Figure \ref{fig:sub3}) and then assess whether the budget is heading in the right direction. As the agent proceeds, she follows along with the status of the steps as they are executed (Figure \ref{fig:sub4}). Once the agent is done, Khushi reviews the spreadsheet and concludes that it matches what she wants. She then presses the `Execute Plan' button to continue with the rest of the plan, and updates her actual data in the final spreadsheet to track her expenses.

\subsection{Agent Instructions} \label{instructions}
We instructed the agent to follow a 3-step workflow across the conversation:
\begin{enumerate}[label=(\arabic*), leftmargin=0pt, itemindent=1.75em, labelsep=0.3em, labelwidth=0.4em]
    \item \textbf{Understand all task context}, primarily the current content in the workbook and any conversation history.
    \item If the query seems ambiguous or has multiple valid approaches to implementation, ask the user \textbf{clarifying questions}.
    \item Based on the user's answers and the workbook context, \textbf{create a step-wise plan} to execute the task for users to review.
\end{enumerate}

This workflow is flexible and interactive, based around the user's actions. For example, if users prompt the agent to make changes to the plan, and these changes seem to have multiple interpretations, the agent may again ask clarifying questions before updating the plan.

To ensure the clarifying questions stage is interactive without overwhelming users, we took inspiration from prior work.
TableTalk \cite{tabletalk} scaffolds spreadsheet development by taking feedback from users at regular stages, and based on this feedback, presents previews of its intended edits to users for review. Generalizing this workflow, we designed Plan Mode to ask users for feedback through clarifying questions, and then present the plan for review (which can possibly lead to a second stage of clarifying questions, based on the user's feedback).
We operationalized this by instructing the agent \textit{not} to repeat any clarifying questions to the user, even if they are unanswered -- if required, the model then makes an assumption about this question, which is surfaced to the user in the plan. This way, users can answer questions which are important to them, without feeling annoyed or overwhelmed by answering questions that may not be relevant to them. Unlike the Plan Modes of tools such as Claude Code and Cursor that require users to answer all clarifying questions through multiple-choice UI cards, our prototype takes responses to clarifying questions through the chat.

Finally, we instruct the agent to restrict the generated plan to 7 steps and one sentence per step, keeping the plan easy to read and edit. 
We also instruct the agent not to include low-level implementation details in the plan, enabling the user and the agent to collaborate on high-level requirements and design decisions for the task in Plan Mode.

\subsection{Design Features} \label{design}
In this subsection, we lay out the design features for our prototype along with the rationale behind them. Some of these features were inspired by those of Plan Modes for commercial coding tools; we lay out the exact relation between our design features and those of such tools in 
Appendix~\ref{app:design_comparison}.
\begin{enumerate}[label=(\arabic*), leftmargin=0pt, itemindent=1.75em, labelsep=0.3em, labelwidth=0.4em]

\item \textbf{Manipulable persistent plan artifact}: Upon creating a plan, our prototype displays this plan in a separate pane of the chat (Figure \ref{fig:sub2}, top). 
This artifact is persistent through planning and execution of the plan, allowing users to refer to it at any point of the process. Further, users can also edit the plan directly -- they can add, remove, reorder, and change the content of steps in the plan, 
allowing for an easy way to update requirements, especially for more minor changes. Prior work has also  observed usability benefits associated with such editable plan representations \cite{cocoa}.
    However, while some tools such as Cursor and VSCode have persistent plans visible to users, most do not allow users to edit these plans in-place through direct manipulation. 
    \item \textbf{Partial Execution}: To ensure that participants remain in control of the agent and the scope of its edits, we added a button that allows users to partially execute the plan (Figure \ref{fig:sub3}, top right). This allows users to iterate at different stages within the task rather than being forced to iterate on a complete task artifact or prematurely pause the agent's execution. Prior work has also found that partial execution within the interactive planning process can improve usability and efficiency of the tool \cite{dango, cocoa}.
    While complex programming tasks may only produce well-formed outputs once all pieces of the code are in place, participants can use intermediate outputs for complex spreadsheet tasks to interpret whether their output is heading in the right direction.
    At the time of writing (June 2026), most popular tools with Plan Modes do not have such a feature.
    \item \textbf{Dynamic Plan Status Updates}: When executing the plan, the agent dynamically updates the status of each step in the plan (Pending $\rightarrow$ In Progress $\rightarrow$ Done) as it is executed (Figure \ref{fig:sub3}), enabling users to track the current state of execution. Tools with persistent plans such as Cursor often have a similar feature, though usually with binary statuses (done/not done). As the agent often makes multiple spreadsheet edits for a single step, the `in-progress' status 
    indicates to participants why certain spreadsheet elements might appear incomplete while the agent works through a particular step of the plan. 
    \item \textbf{Restrictions to edits}: Plan Mode has read-only access to the user's workbook. This allows the user to collaborate with the agent on a plan without worrying about any unexpected edits while building the plan. 
    \item \textbf{User-controlled explicit mode switching}: To ensure the user is aware of and in full control of the current mode the agent is in, we include a prominent button in the chat bar of the tool (Figure \ref{fig:sub1}, bottom right). Users can click on this to toggle the mode on and off. Further, the agent only switches from Plan Mode to execution mode when users click one of the `Execute Plan' buttons (Figure \ref{fig:sub2}, top right).

\end{enumerate}

\subsection{Tool Implementation} \label{implementation}
We built our prototype for Plan Mode as an add-on feature to a high-fidelity internal research prototype (hereafter referred to as the `Act agent') of Excel Copilot \cite{excel_copilot}, an agentic AI tool that can autonomously perform complex tasks within Excel. The Plan Mode prototype served as a way for us to iterate on a viable implementation of a Plan Mode feature for Excel Copilot and to gather feedback on the potential benefits and drawbacks of such a feature. 
Plan Mode has now been released for production within Excel Copilot\cite{excel_copilot}.

The Act agent uses a chat-based interface present in a pane to the right of a Microsoft Excel window; users can enter their queries into a chat box following which the agent performs some internal reasoning (surfaced to users), runs commands to read from and make edits to the spreadsheet, and eventually responds to the query with a summary of its findings and changes made to the spreadsheet. The Act agent contains detailed instructions for reading and writing to Excel spreadsheets, and is instructed to execute the user's tasks immediately and autonomously. It does not have the capability to create or store plans for task execution.
The Plan Mode feature uses the same interface with the addition of the Plan pane at the top of the chat (Figure \ref{fig:sub2}, top), and the `Plan Mode ON/OFF' button at the right side of the chat input box (Figure \ref{fig:sub1}, bottom right).

In early experiments, we observed that providing a single agent with instructions for both planning and task execution led to the agent sometimes using tools it was not meant to use, such as trying to edit the spreadsheet in Plan Mode or edit the plan during execution. We thus decided to build a separate Plan agent for Plan Mode to cleanly separate the instructions for these two phases. When users interact with the tool in Plan Mode, they interact with the Plan agent, and when they switch to execution, the Act agent takes over.
This Plan agent does not have access to any instructions or tools for editing the spreadsheet, but has access to an extra tool to render the plan in the Plan pane. Similarly, the Act agent does not have access to the instructions for plan creation or to the tool for rendering the plan, but has read-only access to the plan along with a tool to update the status of the steps of the plan in the UI. Both the Plan agent and the Act agent share the same conversation history so that they have the same context. We used the LLM Claude Sonnet 4.5 for all sessions in our study.

\section{Methodology} \label{methods}
Plan Mode involves a fundamentally distinct interaction paradigm for human-AI collaborations; rather than asking and resolving queries one-by-one, users now work with the agent in a structured multi-turn workflow. Yet, the technical capabilities of the agent in terms of spreadsheet actions and base LLM remain the same across modes. Our evaluation thus assesses how these differences and similarities manifest in practical usage through the following research questions:
\begin{enumerate}[label=\textbf{RQ\arabic*:}, leftmargin=*, align=left]
    \item How does interactive planning affect how end-user programmers \textbf{exchange information} with AI tools on open-ended spreadsheet tasks?
    \item How does interactive planning affect the \textbf{final outputs} created by end users when working with AI tools on open-ended spreadsheet tasks?
    \item How does interactive planning affect end users' \textbf{perceived experiences} when working with AI tools on open-ended spreadsheet tasks?
\end{enumerate}

To do so, we conducted a controlled mixed-methods within-subjects user study (N = 24). In each session, participants completed one task each with Plan Mode and the Act agent (hereafter referred to as `Act Mode').
All sessions were conducted in February 2026.

\begin{table}[t]
\caption{Participant Demographics}
\centering
\footnotesize
\begin{tabular}{@{}p{1.2cm}p{6.8cm}@{}}
\toprule
\textbf{Dimension} & \textbf{Details} \\
\midrule

\rowcolor{gray!5}Age &
18--24: 1; 25--34: 9; 35--44: 9; 45--54: 2; 55--64: 3 \\

\arrayrulecolor{gray!50}\cmidrule{1-2}\arrayrulecolor{black}

\rowcolor{gray!5}Gender &
Men: 13; Women: 11; Additional genders: 0 \\

\arrayrulecolor{gray!50}\cmidrule{1-2}\arrayrulecolor{black}

\rowcolor{gray!5}Ethnicity &
White/Caucasian: 14; Asian or Pacific Islander: 5; Black or African American: 4; Hispanic or Latino: 1 \\

\arrayrulecolor{gray!50}\cmidrule{1-2}\arrayrulecolor{black}

\rowcolor{gray!5}Region &
US: 13; UK: 7; Canada: 2; France: 1; Australia: 1 \\

\arrayrulecolor{gray!50}\cmidrule{1-2}\arrayrulecolor{black}

\rowcolor{gray!5}Industry &
Technology: 8; Healthcare: 4; Finance: 4; Retail: 2; Marketing/Advertising: 1; Logistics/Supply Chain: 1; Other: 4 \\

\bottomrule
\end{tabular}
\label{tab:demographics}
\vspace{-1em}
\end{table}

\subsection{Participants}
Participant demographics are described in Table \ref{tab:demographics}. We recruited participants through an online user study recruitment platform (PlaybookUX, \url{playbookux.com}). Through a screener survey (available in supplementary material), we ensured that our participants used spreadsheets at least occasionally (more than once a week) 
and that they were familiar with creating and manipulating spreadsheets using spreadsheet software. We also ensured that participants had experience with using AI tools previously.
Further, we excluded those who were `beginners' with spreadsheet software (that is, those who were not familiar with formulas), so that participants were capable of interpreting the edits made by the tool. 

\subsection{Tasks}
For our study, we aimed to use tasks that are open-ended and broadly subject to personal context and preferences -- since such tasks have multiple valid approaches and solutions, planning can have an important role to play in shaping the final output.
We further used two distinct types of tasks to understand the impact of planning across different tasks: 
\textbf{\Creation} tasks in which participants had to create workbooks from scratch, and open-ended \textbf{\Analysis} tasks in which participants had to perform data analysis on a large dataset based on their subjective preferences to pick final results. These correspond to the ``Creation - Artifact'' and ``Information - analyze'' categories of the knowledge worker LLM use taxonomy by Brachman et al. \cite{brachman2024llmusetaxonomy} and are the activities most directly supported by the Act agent.

We further sought tasks that could be completed in around 15 minutes, and use typical spreadsheet artifacts (such as formulas and charts) without requiring too much technical complexity. Thus, we used the following two \Creation tasks: 1) \Budget: Participants were instructed to create a personal budget, with the objective of creating a useful spreadsheet to track their expenses, and 2) \Schedule: Participants were instructed to create a personal schedule with the objective of using the spreadsheet to manage their time more efficiently.

Similarly, we used the following two \Analysis tasks: 1) \Vacation: Participants were provided with a large dataset with details about vacation destinations (such as things to do, temperature, etc.) as well as details about accommodations in these destinations (price, ratings, etc.), and were asked to plan their next vacation based on the details in the data. 2) \Movie: Participants were provided with a dataset of movies with details such as genres, ratings, popularity, etc. and asked to find a movie for them to watch. We include the task descriptions provided to participants and the datasets for the analysis tasks in the supplementary material.

In each session, participants performed either both \Creation tasks or both \Analysis tasks. We paired tasks within the same type so that each participant's comparisons across different modes were made on comparable tasks, isolating the effect of mode from the effect of task type. We balanced the type of tasks as well as the order of the specific tasks and tools participants used in each session.

\subsection{Protocol}
All user study sessions were conducted by the first author through a video conferencing tool, and all participants worked on their tasks by remotely controlling the study administrator's screen. Each session started with the study administrator giving participants a tutorial of the first tool participants used (Plan Mode/Act Mode), and answering any questions they had about the tool. While Act Mode consisted primarily of a familiar chat-based interface, Plan Mode consisted of potentially unfamiliar UI elements such as the Plan pane (Figure \ref{fig:sub2}, top). Thus, our tutorial for Plan Mode also included a demonstration of the different UI elements of the tool.
Following this, the study administrator described the first task to participants, and instructed them that they had around 15 minutes to work on their task (in practice, we allowed up to 20 minutes for participants to work on their tasks).  We also instructed participants to ``think aloud" during their tasks so as to understand their process and reactions to the tool. 
If participants mentioned that they were satisfied with the workbook before the 20-minute time limit (in practice, this happened in most cases), the study administrator asked them if they would like to make any other changes to the workbook. If participants still indicated that they would not make any changes beyond data entry (such as entering their real expense amounts for the \Budget task), they moved on to the second task, for which the study administrator repeated the same process. 
After both tasks were complete, participants were instructed to fill out a post-study questionnaire.
We asked participants to answer the questionnaire questions based on their experience with the tool rather than their experience working on the task to more accurately measure their preferences for the tool. Further, the study administrator asked participants to walk through their answers to understand the reasons behind their preferences. Each session was scheduled for one hour, and took between 30 and 60 minutes based on how much time participants took to complete their tasks.

\begin{table}[t]
\centering
\footnotesize

\begin{tabular}{p{0.22\columnwidth} p{0.68\columnwidth}}
\toprule
\textbf{Label} & \multicolumn{1}{c}{\textbf{Description}} \\
\midrule

\multicolumn{2}{c}{\textbf{Origin}} \\
\midrule

\rowcolor{gray!10}
{proactive} &
User mentions the requirement proactively, not based on any message, plan, or clarifying question presented by the agent \\

{clarification} &
User mentions the requirement as a reply to a clarification question by the agent \\

\rowcolor{gray!10}
{plan} &
User mentions the requirement after seeing a plan generated by the agent [only for Plan Mode] \\

{refine} &
User mentions the requirement after seeing edits the agent made to the workbook \\

\midrule
\multicolumn{2}{c}{\textbf{Iteration}} \\
\midrule

\rowcolor{gray!10}
{new} &
The first time the requirement is mentioned \\

{follow\mbox{-}up} &
Follow-up on an existing requirement (e.g., deeper or complementary information) \\

\rowcolor{gray!10}
{reiteration} &
The user repeats an existing requirement without any changes \\

{change} &
The user revises an existing requirement \\

\bottomrule
\end{tabular}

\caption{User requirement coding}
\label{tab:req_codebook}
\vspace{-2em}
\end{table}

\subsection{Data Collection and Analysis}

\subsubsection{Conversation Logs}
We collected the conversation logs for participants' interactions with both tools. We used this data to aggregate general information about each task, such as the number of turns, duration, and number of tokens created by the LLM (approximate calculation by dividing the generated reasoning, messages, and tool call arguments by 4).

We also used this data to code the unique requirements participants expressed across their prompts. For \Creation tasks, we consider a requirement to be a new \textit{type} of information mentioned by the user (such as expense categories, a specific chart, etc.) while for \Analysis tasks, we consider a requirement to be any {filtering criterion or sorting criterion} for a particular column in the dataset. We also coded the origin of these requirements and the iteration type of the requirement, as described in Table \ref{tab:req_codebook}.  

While the codebook for requirement origin was based on the interaction mechanisms of Plan Mode and Act Mode, the codebook for requirement iteration was initially developed by the first author through open coding of two conversation logs, and refined to form the final version after coding two more such logs. To validate these codebooks, two authors independently coded a stratified sample of 12.5\% of the data (6 out of 48 conversations, covering all tasks and treatments). Disagreements after the first round of coding were resolved by updating the codebook based on discussion between the two raters \cite{mcdonald2019reliability}, followed by a second round of coding. We used Cohen's Kappa to measure inter-rater reliability, yielding scores of 
 $\kappa$ = 0.93 (quadratic-weighted) for extracting requirements from raw conversation transcripts, $\kappa$ = 0.62 for the requirements iteration coding and $\kappa$ = 0.85 for the requirements origin coding, which indicate substantial to near-perfect agreement~\cite{cohen-kappa}.

\subsubsection{Spreadsheets Created}
We collected all spreadsheets created by participants and analyzed the \textit{features} present across these spreadsheets, applying the following criteria:
\paragraph{\Creation tasks} We applied a hierarchical coding system to analyze the contents of each workbook, noting the number of top-level spreadsheet artifacts (sheets, tables, and charts) present across these workbooks as well as the unique features present in these artifacts. Each feature consists of one primary code (e.g., `expenses') associated with one or more secondary codes (e.g., `taxes'). We include the complete feature codebook in the supplementary material.
For each task, the codebook was initially developed through an open coding of two workbooks generated by participants. It was then refactored to be hierarchical based on repeating themes across the open coding, and refined by re-coding the same two workbooks along with one additional workbook. 

\paragraph{\Analysis tasks} We note each information column used by the participant in picking their final results, along with whether these columns were associated with any filtering criteria, sorting criteria, or both. Often, the agent created a score using information from several columns -- in these cases, we counted all of these columns as features. 

\begin{table}[t]
\centering
\footnotesize
\renewcommand{\arraystretch}{1}

\begin{tabular}{p{0.015\columnwidth} p{0.18\columnwidth} p{0.64\columnwidth}}
\toprule
\textbf{\#} & \textbf{Dimension} & \textbf{Question(s)} \\
\midrule

\multicolumn{3}{c}{\cellcolor{gray!10}\textbf{Creativity Support Questions}} \\
\midrule

\rowcolor{gray!10}
1 &
\multirow{3}{*}{Enjoyment} &
``I would be happy using this tool more on a regular basis.'' \\

2 & &
``I enjoyed using the tool more.'' \\

\midrule

\rowcolor{gray!10}
3 &
\multirow{6}{*}{Exploration} &
``It was easier for me to explore many different ideas, options, designs, or outcomes, using this tool.'' \\

4 & &
``This tool was more helpful in allowing me to track different ideas, outcomes, or possibilities.'' \\

\midrule

\rowcolor{gray!10}
5 &
\multirow{3}{*}{Expressiveness} &
``I was able to be more creative while doing the activity inside this system or tool.'' \\

6 & &
``This tool allowed me to be more expressive.'' \\

\midrule

\rowcolor{gray!10}
7 &
\multirow{2}{*}{Attention*} &
``My attention was more focused on the task.'' \\

8 & &
``My attention was more focused on the tool.'' \\

\midrule

\rowcolor{gray!10}
9 &
\multirow{5}{*}{Worth} &
``I was more satisfied with what I got out of this tool.'' \\

10 & &
``What I was able to produce was more worth the effort I had to exert to produce it with this tool.'' \\

\midrule

\multicolumn{3}{c}{\cellcolor{gray!10}\textbf{Human-Machine Collaboration Questions}} \\
\midrule

\rowcolor{gray!10}
11 & Communication &
``I was able to more effectively communicate what I wanted to the system.'' \\

12 & Alignment &
``I was able to steer this tool better towards output that was aligned with my goals.'' \\

\rowcolor{gray!10}
13 & Agency &
``At times, I felt that this tool was steering me towards its own goals.'' \\

14 & Partnership &
``At times, it felt like this system and I were collaborating as equals.'' \\

\midrule

\multicolumn{3}{c}{\cellcolor{gray!10}\textbf{Exploratory Questions}} \\
\midrule

\rowcolor{gray!10}
15 & Contribution* &
``I am creatively responsible for this spreadsheet'' vs ``The agent is creatively responsible for this spreadsheet'' \\

16 & Satisfaction &
``I'm very unsatisfied with the spreadsheet'' vs ``I'm very satisfied with the spreadsheet.'' \\

\rowcolor{gray!10}
17 & Surprise &
``The spreadsheet was what I was aiming for'' vs ``The spreadsheet outcome was unexpected.'' \\

18 & Novelty &
``The spreadsheet is very typical'' vs ``The spreadsheet is very novel.'' \\

\rowcolor{gray!10}
19 & Capability* &
``I could have created the spreadsheet by myself'' vs ``I could not have created the spreadsheet without using this tool'' \\

\bottomrule
\end{tabular}

\caption{\centering Post-study questionnaire used in our study, based on \cite{micsi} (* indicates dimensions modified for our study)}
\label{tab:questionnaire}
\vspace{-1.5em}
\end{table}

\subsubsection{Post-study questionnaire} We adapted the Mixed-Initiative Creativity Support Scale \cite{micsi}. The first 10 questions of this scale focus on creativity support along 5 dimensions; the next 4 capture Human-Machine Collaboration; and the last 4 use paired statements to elicit perceptions of the outputs. 

To compare experiences between modes rather than capture participants' experiences with agentic spreadsheet tools in general, we changed the phrasing of the first 14 questions of the survey to be comparative across tools (for example, \textit{``I enjoyed using this tool more"} rather than \textit{``I enjoyed using this tool"}), and used a preference-based 5-point scale to collect responses [\textit{`Act Mode', `Somewhat Act Mode', `Neutral', `Somewhat Plan Mode', `Plan Mode'}]. For the remaining questions, we asked participants to select the statement that most closely matched their experience for each tool. We also adapted the `Immersiveness' dimension to a more relevant `Attention' dimension, modified the phrasing of the `Contribution' statement, and added a new dimension for `Capability' (details in supplementary material).
The final questionnaire used is described in Table \ref{tab:questionnaire}. 

\subsubsection{Session Recordings} We used video, audio and screen recordings of the sessions to note whether and how participants interacted with the Plan Mode UI, and to code participant utterances during the think-aloud portions of the study.

\subsubsection{Limitations} \label{limitations}
The small sample size (N=24) of our study limits the generalizability of our results; we accept this limitation since it allowed us to conduct in-depth qualitative coding of participant requirements and workbooks, which would be difficult to scale to larger samples. Another limitation of our study is that almost all participants in our study had not used \textit{agentic} tools for spreadsheet programming previously, which may have affected participants' experiences with the tool. We also restricted our tasks to those which could be completed in a short time frame (15 minutes), which may have limited our ability to observe the differences between the two modes. This is evident in the fact that while the comparative measurement instrument (Figure \ref{fig:questionnaire_pt1}) shows differences, the non-comparative instrument (Table \ref{tab:questionnaire_pt2}) shows near-identical ratings across tools.

Further, the fact that Plan Mode had unfamiliar UI elements as compared to Act Mode and that our tutorial for Plan Mode covered UI elements while the tutorial for Act Mode did not is a confound that may have affected participants' perceptions of the tool. Additionally, while we report descriptive statistics related to spreadsheet artifacts and total number of requirements, the time-capped nature of each task may have affected these numbers, especially for the participants who did not fully complete the task (though the number of such participants was low).
Finally, we acknowledge our positionality as researchers at Microsoft, a technology company that develops AI tools, and any unconscious biases that this positionality might engender in the interpretation of our data.

\section{Results}

\begin{table}[t]
\centering
\begin{tabular}{lcccc}
\toprule
 & Proactive & Clarification & Plan & Refine \\
\midrule
Plan Mode & 3.1 & 4.1 & 0.7 & 1.3 \\
Act Mode  & 4.3 & 0.2 & - & 3.1 \\
\bottomrule
\end{tabular}
\caption{Mean requirements by origin across modes}
\label{tab:reason_mode}
\vspace{-0.7em}
\end{table}

\begin{table}[t]
\centering
\begin{tabular}{lcccc}
\toprule
 & New & Follow-up & Reiteration & Change \\
\midrule
Plan Mode & 6.5 & 2.0 & 0.1 & 0.6 \\
Act Mode  & 6.4 & 0.7 & 0.1 & 0.4 \\
\bottomrule
\end{tabular}
\caption{Mean requirements by iteration across modes}
\label{tab:repetition_mode}
\vspace{-2em}
\end{table}

\begin{table}[t]
\centering
\begin{tabular}{lcccc}
\toprule
 & Proactive & Clarification & Plan & Refine \\
\midrule
Plan Mode & 46\% & 37\% & 7\% & 10\% \\
Act Mode  & 64\% & 1\% & - & 35\% \\
\bottomrule
\end{tabular}
\caption{Origins of \textbf{new} requirements across modes}
\label{tab:new_reason}
\vspace{-1.3em}
\end{table}

\subsection{RQ1: Interaction Patterns}
RQ1 asks how interactive planning affects how end-user programmers exchange information with AI tools on open-ended spreadsheet tasks. We explore these patterns across three sub-aspects: the differences in how participants expressed their requirements to the agent across modes (\S \ref{reqs}), whether and how different information exchange patterns changed how participants iterated on their spreadsheets (\S \ref{complex}), and whether participants used the planning UI affordances to exchange information (\S \ref{UI}).

\subsubsection{Requirements Elicitation} \label{reqs}

Our results suggest that participants using Plan Mode expressed both more requirements and more detailed requirements before the agent made edits to the workbook, a shift largely driven by the presence of clarifying questions.

Participants expressed a similar number of unique requirements across modes (Table \ref{tab:repetition_mode}, ``New”), but arrived at them through different interaction patterns. Participants were highly responsive to clarifying questions in Plan Mode, directly answering 88 of the 99 questions asked by the agent. As a result, participants in Plan Mode expressed 44\% of their requirements in response to clarifying questions, while only 14\% emerged after the agent had already edited the spreadsheet. In contrast, participants in Act Mode expressed over 41\% of their requirements as refinements to existing spreadsheet edits (Table \ref{tab:reason_mode}).

This shift in requirements from refinements to clarifying questions in Plan Mode is also reflected in the origin of new requirements expressed by participants. When using Plan Mode, 10\% of participants' new requirements came from refinements, while 37\% came from clarifying questions. This was flipped for participants using Act Mode, for whom 35\% of new requirements stemmed from refinement requests.

Beyond eliciting new requirements, clarifying questions also helped participants dive into more depth on these requirements.
When using Plan Mode, participants provided more `follow-ups' (updates and complementary information) to their requirements (Table \ref{tab:repetition_mode}, `Follow-up'), and the majority (82\%) of follow-ups on requirements in Plan Mode came from responses to clarifying questions. Thus, clarifying questions not only surfaced new requirements, but also drew out additional detail on existing ones.

\subsubsection{Refinement Patterns} \label{complex}
Our results suggest that spending time on upfront planning enabled participants to build spreadsheets more aligned with their goals.
Participants using Plan Mode used more turns (5.3 vs 3.6, on average) and took more time (10.4 vs 8.8 minutes, on average
) across all tasks in our study.
At the same time, we found that participants spent more turns refining the spreadsheet when working with Act Mode (2.2 vs 1.4 on average)
and that they often did not iterate on plans at all -- only 10 out of 24 participants made any changes to the first proposed plan through chat messages or the UI. 

This reduction in refinement also results in a reduction in cost. Despite the fact that participants expressed a similar number of total requirements for both treatments, the approximate number of tokens used in the LLM's reasoning, messages, and spreadsheet edits was much lower with Plan Mode than with Act Mode across all tasks (11.0k vs. 17.0k on average).

Our results also suggest that participants found the \Budget and \Vacation tasks to be lengthier than the \Schedule and \Movie tasks respectively -- on average, they used more conversation turns (4.9 vs 4.0) and spent more time (11.9 vs 7.3 minutes) on these tasks. 

\subsubsection{Using the UI} \label{UI}
We found that while participants did not tend to use the UI features of Plan Mode, they often still appreciated the control these features offer. 6 out of 24 participants chose to partially execute the plan by using the `Execute Step 1-x' button (see Figure \ref{fig:sub3}, top right). Further, only 2 out of 24 participants edited the plan representation in the Plan Mode UI. Yet, 14 out of 24 participants spoke positively about the ability to edit/rearrange the steps of the plan as a way for them to be able to control or steer the agent better. As P8 said, \textit{``I could see, you know, it's like creating a game plan, and if I'm missing something from my game plan, I can add to my game plan.''}

\subsection{RQ2: Differences in generated spreadsheets}
Overall, we found little difference in the distribution of features in participants' spreadsheets across modes, indicating that working with Plan Mode does not lead to the creation of more diverse or personalized spreadsheets. 
However, we did find some task-wise differences, with the lengthier tasks generally using/containing more features. 

\subsubsection{\Analysis Tasks}
Participants used data from more columns for the \Vacation task (median 7 unique columns used for filtering and/or sorting) than for the \Movie task (median 4 unique columns used for filtering and/or sorting). We did not find any notable differences in the number of columns used for filtering and sorting across modes (median: 4.5 in Plan Mode vs 5 in Act Mode). We also found that participants used similar columns for their analysis in each task across both modes (details in 
Appendix \ref{app:additional_figures}).

\subsubsection{\Creation Tasks}
For the creation tasks, we found that participants using Act Mode tended to create more spreadsheet artifacts than those using Plan Mode (Table \ref{tab:artifact_counts}), particularly for the \Budget task. However, the spreadsheets created with Act Mode contained only a moderately larger number of total features across both tasks (median values for Plan vs Act: 9 vs 10.5 for the \Budget task; 3 vs 4.5 for the \Schedule task). Many of the extra features for the \Schedule task in Act Mode correspond to meta-information about the spreadsheet and instructions rather than more types of data. Thus, overall, the spreadsheets generated contained similar features across modes despite Act Mode using more spreadsheet artifacts.

\begin{table}[t]
\centering
\begin{tabular}{llccc}
\toprule
Task & Mode & Sheets & Tables & Charts \\
\midrule
\multirow{2}{*}{Budget}
  & Plan Mode & 1.5 & 4.5  & 2 \\
  & Act Mode  & 2.5 & 10.5 & 2 \\
\midrule
\multirow{2}{*}{Schedule}
  & Plan Mode & 1.5 & 3 & 0 \\
  & Act Mode  & 1 & 4.5 & 0 \\
\bottomrule
\end{tabular}
\caption{\centering Median number of sheets, tables, and charts created by mode and task}
\label{tab:artifact_counts}
\vspace{-1.5em}
\end{table}

We also found that participants created more artifacts when working on the \Budget task (48 sheets, 138 tables and 30 charts) as compared to when working on the \Schedule task (23 sheets, 67 tables and 7 charts), further indicating that participants found the \Budget task to be lengthier.

\begin{figure}[t]
    \centering

    \begin{subfigure}{\linewidth}
        \centering
        \includegraphics[width=1.05\linewidth]{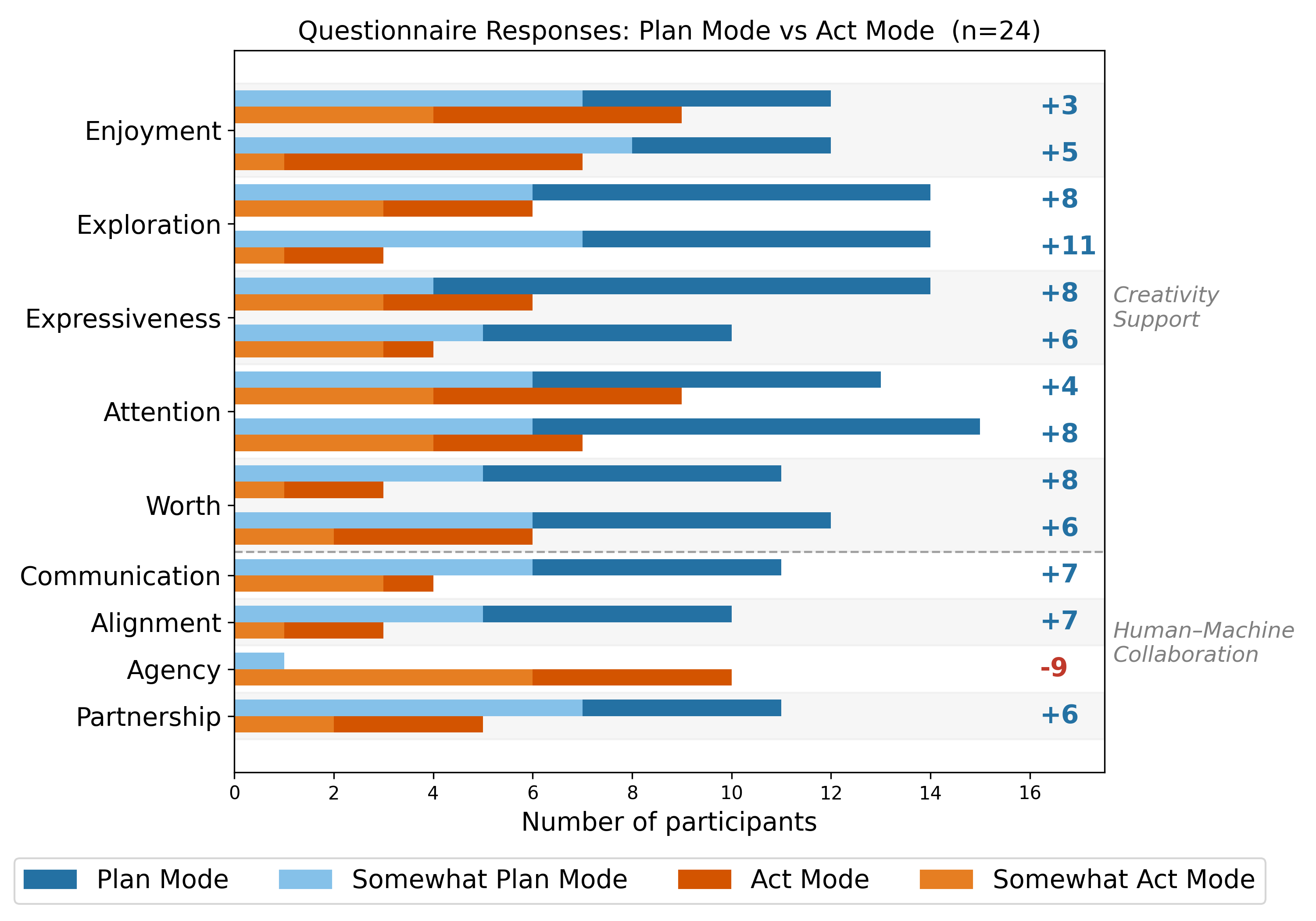}
        \caption{Questionnaire responses (Part 1) (Neutral responses not shown)}
        \label{fig:questionnaire_pt1}
    \end{subfigure}

    \vspace{0.75em}

    \begin{subfigure}{\linewidth}
        \centering
        \footnotesize
        \renewcommand{\arraystretch}{1}

        \begin{tabular}{p{0.14\columnwidth} p{0.6\columnwidth} cc}
        \toprule
        \textbf{Category} & \textbf{Statement} & \textbf{Plan} & \textbf{Act} \\
        \midrule

        \rowcolor{gray!10}
        \multirow{2}{*}{Contribution}
        & The agent is creatively responsible for this spreadsheet
        & 8 & 10 \\

        & I am creatively responsible for this spreadsheet
        & 16 & 14 \\

        \midrule

        \rowcolor{gray!10}
        \multirow{2}{*}{Satisfaction}
        & I'm very satisfied with the spreadsheet
        & 22 & 23 \\

        & I'm very unsatisfied with the spreadsheet
        & 2 & 1 \\

        \midrule

        \rowcolor{gray!10}
        \multirow{2}{*}{Surprise}
        & The spreadsheet was what I was aiming for
        & 19 & 20 \\

        & The spreadsheet outcome was unexpected
        & 5 & 4 \\

        \midrule

        \rowcolor{gray!10}
        \multirow{2}{*}{Novelty}
        & The spreadsheet is very typical
        & 12 & 14 \\

        & The spreadsheet is very novel
        & 12 & 10 \\

        \midrule

        \rowcolor{gray!10}
        \multirow{2}{*}{Capability}
        & I could have created this spreadsheet by myself
        & 17 & 18 \\

        & I could not have created this spreadsheet without AI
        & 7 & 6 \\

        \bottomrule
        \end{tabular}
        \caption{Questionnaire responses (Part 2)}
        \label{tab:questionnaire_pt2}
    \end{subfigure}

    \caption{Questionnaire responses by participants}
    \label{fig:questionnaire_combined}
    \vspace{-1em}
\end{figure}

\subsection{RQ3: Perceived experiences}

We found that participants preferred Plan Mode across almost all dimensions of creativity support and human-machine collaboration and had similar perceptions of the spreadsheets they created (Table \ref{tab:questionnaire_pt2}), though this relationship was moderated by both the types of tasks participants worked on as well as their own interaction styles. Figure \ref{fig:questionnaire_combined} shows a summary of participants' responses on our post-study questionnaire. Participants preferred Plan Mode over Act Mode across all five dimensions of creativity support (each dimension corresponds to two questions) and all four questions on effective collaboration. Participants answered Act Mode more frequently on the `Agency' question, since this asks about the extent to which the tool steers the user towards its own goals. The only exception is the second question of the Attention scale -- participants indicated that working with Plan Mode required more of their attention on the tool, which may be due to the extra UI elements present in Plan Mode as compared to Act Mode. Table \ref{tab:questionnaire_pt2} also indicates that despite participants preferring Plan Mode for creativity support, they feel a similar amount of creative responsibility and satisfaction with the output when working with Plan and Act Mode. These preferences were moderated by several factors:

\begin{figure}[t]
    \centering
    \includegraphics[width=1.1\linewidth]{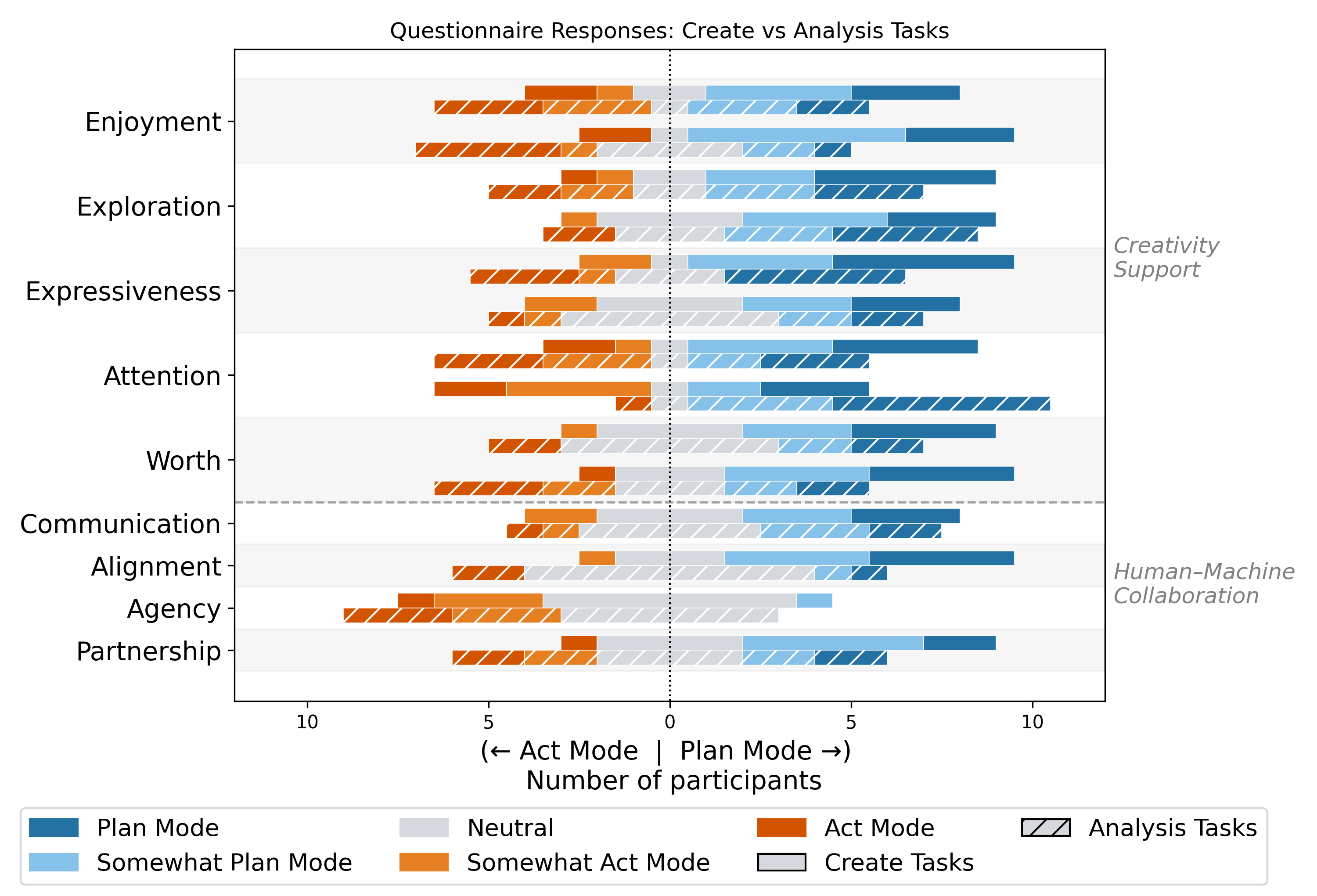}
    \caption{Participant preferences based on task type}
    \label{fig:questionnaire_task_type}
     \vspace{-1em}
\end{figure}

\begin{figure}[t]
    \centering
    \includegraphics[width=1.1\linewidth]{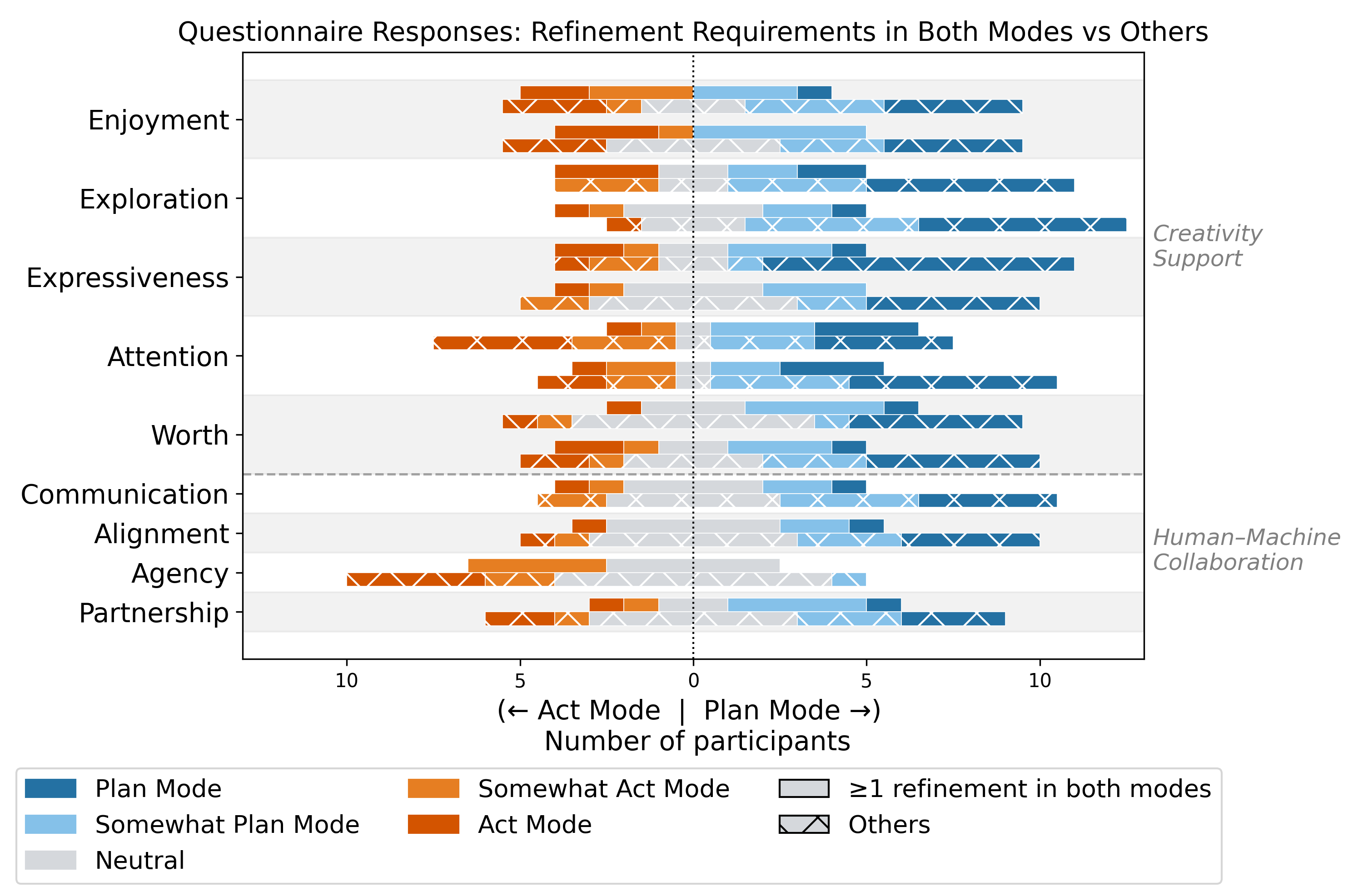}
    \caption{Participant preferences based on refinements}
\label{fig:plan_mode_preference_by_refinement_requirements}
\vspace{-1.5em}
\end{figure}

\begin{enumerate}[label=(\arabic*), leftmargin=0pt, itemindent=1.75em, labelsep=0.3em, labelwidth=0.4em]
    \item \textbf{Task Type:} Participants preferred Plan Mode more for \Creation tasks across all dimensions, while they were more ambivalent for \Analysis tasks, as highlighted in Figure \ref{fig:questionnaire_task_type}. The contrasts across the dimensions of enjoyment, worth, alignment, and partnership are particularly stark -- those working on \Analysis tasks were neutral or leaned towards Act Mode for these dimensions, while those working on \Creation tasks strongly preferred Plan Mode. These specific dimensions indicate that participants working on \Creation tasks felt more like they were interacting with a collaborative partner when working with Plan Mode.
    Further, those working on \Analysis tasks felt that they had to pay more attention to the tool rather than the task when working with Plan Mode (unlike for \Creation tasks), highlighting that participants working on \Creation tasks found the experience smoother and more collaborative with Plan Mode.
    \item \textbf{Task Length:} Participants preferred Plan Mode more for the tasks they spent more time and turns on (\Budget and \Vacation, \S \ref{complex}). These participants particularly preferred Plan Mode across the dimensions of agency, exploration, worth and partnership, while other participants were more neutral on these dimensions. This indicates that Plan Mode provided participants with more control and a more effective collaboration
    when working on lengthy tasks.
    
    \item \textbf{Number of upfront requirements:} Participants who provided more than the mean (3.7) number of requirements upfront in both their tasks (n = 9) enjoyed Act Mode more than Plan Mode and felt they could explore ideas and outcomes more easily with it. This suggests Plan Mode was more helpful for those who were less effective at one-shot prompting.
    \item \textbf{Number of refinements:} While participants who expressed requirements as refinements tended to be more neutral,  others strongly preferred Plan Mode across all creativity support dimensions, as well as the communication dimension. Figure \ref{fig:plan_mode_preference_by_refinement_requirements} shows the questionnaire responses for those participants (n=9) who expressed at least one requirement as a refinement across both modes, as compared to the other participants (n = 15). Thus, participants who followed the pattern of iterative refinement found that Plan Mode did not help them in being creative. 7 participants explicitly mentioned that they liked Act Mode since it facilitated iterative refinement in a way that was easier or quicker than with Plan Mode -- it allowed users to view and update outputs in \textit{``real time''} (P2). These preferences seemed to originate from the process of planning itself not matching participants' workflow. For example, P16 mentioned that \textit{``Plan Mode seemed like you really had to plan, and if I'm gonna do that in an AI, I might as well just do that myself... Act Mode is more natural and freestyle. You can just change your mind on the fly.''}
\end{enumerate}

\section{Discussion}
Our results suggest that using Plan Mode changed how participants collaborated with the agent on their tasks in terms of requirements and refinements and further resulted in a better perception of this collaboration, despite the final outputs created being similar.
In this section, we discuss potential design interventions that might allow Plan Mode to engender higher quality outputs for users (\S \ref{d-design}), why participants may have preferred Plan Mode despite the similarity in outputs (\S \ref{d-control}), and future research directions for the design and personalization of Plan Mode for end-user programming (\S \ref{d-entry}).

\subsection{Design Implications for Plan Mode} \label{d-design}
Though participants expressed a similar number of unique requirements across modes, it is interesting to note where these new requirements came from -- while all requirements when using Act Mode came from participants, 37\% of requirements when using Plan Mode emerged from a shared origin (clarifying questions, asked by the agent and answered by the user) (Table \ref{tab:new_reason}). However, this new source of requirements did not manifest in the final outputs,
indicating that the agent may not actually be contributing anything particularly new -- it might be asking questions that users would answer anyway (perhaps during the refinement stage) or that the tool in Act Mode might implement anyway. This may be exacerbated by automation bias \cite{automation_bias} -- since participants answered most clarifying questions and usually did not iterate on the plan (Section \ref{complex}), the outputs produced may be strongly influenced by the dimensions the agent thinks are important (expressed through clarifying questions) and follow an execution path the agent thinks is best (expressed through the plan which users did not modify). These homogeneous results might also indicate that agents across both Plan Mode and Act Mode may be reducing, rather than increasing, users' creativity \cite{lowdiversity}. 

Thus, the paradigm of clarifying questions should be modified to augment users' thinking  \cite{challenge,toolsforthought}. Since spreadsheet users can struggle to discover new features in spreadsheet software \cite{discover_features}, augmenting their thinking also provides an opportunity for users to do so in an environment scaffolded by an agent. For example, Drosos et al. \cite{promptions} report that dynamically generated form-based clarifying questions can help participants reflect on their own assumptions, enabling them to explore alternative solutions. Similarly, Danry et al.\cite{dont_tell_ask} report that framing AI-generated explanations as questions can increase critical thinking; such a paradigm can fit naturally into a Plan Mode.
Future work can explore how other such interfaces, which apply strategies like design friction \cite{friction} and cognitive forcing functions \cite{forcing} to promote critical thinking, can augment users' thinking and creativity.

\subsection{Control Without Action as Design Value} \label{d-control}
Despite the overall similarities in outputs and requirements between Plan Mode and Act Mode, participants indicated a preference for Plan Mode in the post-study questionnaire (Figure \ref{fig:questionnaire_pt1}). We observe a similar dichotomy in the usage of the UI features of Plan Mode -- despite almost never using these features, participants indicated that the UI provided them with a sense of control over the agent (Section \ref{UI}).  Kazemitabaar et al. \cite{steering_and_verification} report a similar finding in the computational notebook domain -- though steering mechanisms for AI tools led to no difference in task performance metrics, participants in their study reported greater feelings of control with these mechanisms. 
This visual tracking and manual control over the agent's actions may serve as a way for users to overcome the ``gulf of envisioning" with generative AI, allowing users to better operationalize their goals when collaborating with an agent. 
Thus, such UI features, even when not used, provide design value in and of themselves -- though they may not increase performance metrics, they may still improve the quality of the human-AI collaboration.

\subsection{Tasks, Interaction Styles, and Mixed-Initiative Entry} \label{d-entry}
Most implementations of Plan Modes leave entry and exit completely up to the user. Yet, end-user programmers might not be equipped to know \textit{when} to utilize Plan Mode, since they may not have the skills to understand which tasks are complex and may benefit from planning \cite{sarkar2022likeprogramartificialintelligence, euse}. 
Thus, an important future direction is to investigate mixed-initiative paradigms for triggering Plan Mode.

Our results suggest that different types of spreadsheet programmers react to Plan Mode differently -- participants who specified many requirements upfront and those who iterated on workbooks across both modes tended to enjoy Act Mode more than Plan Mode. These patterns resemble those of a selective information processing style \cite{gendermag}, in which end-users tend to apply a more ``depth-first" approach to problem-solving. This is in contrast to the comprehensive information processing style, in which end-users gather all information to form a complete understanding of the problem before proceeding -- a more plan-like approach. Further, we found that participants preferred Plan Mode more when working on \Creation tasks, which, while similarly open-ended as the \Analysis tasks, required more comprehensive spreadsheet edits.
Future work should thus explore how agents can infer and utilize users' information processing style along with the context of their task to scaffold the integration of Plan Mode into their workflow. Such a paradigm would support end-user programmers in extracting the best out of their agents without imposing the additional metacognitive overhead of adjusting to different modes or features.

\section{Conclusion}
In this paper, we presented a prototype of a Plan Mode for spreadsheet programming agents that allows users to plan their task with AI before they execute the task. In a within-subjects user study (N=24), we found that using Plan Mode affected how participants exchanged information with the AI and improved their perception of the collaboration with the AI. 
Our findings have since informed the design and deployment of the Plan Mode feature of Excel Copilot in Microsoft Excel, where we continue to observe positive feedback on the impact of planning for human-AI collaborations.

\section{Acknowledgements}
We would like to thank the whole feature crew of Plan Mode for Excel Copilot for their support throughout this project. We especially want to thank Avani Reddy for their valuable insights and feedback throughout this project, Arturo Goicochea for their insights on the design of the Plan Mode prototype, as well as San Lee and Saloni Gupta for their help in implementing Plan Mode.

\bibliographystyle{ieeetr}
\bibliography{references}%

\appendix

\subsection{Relation of Design Features to Existing Plan Modes}
\label{app:design_comparison}
 
Table~\ref{tab:design_comparison} summarizes how each of our prototype's design features relates to the Plan Modes of four coding agents -- Claude Code~\cite{claude_code}, VSCode Agent Mode~\cite{vscode_plan_agent}, Cursor~\cite{cursor_plan_mode}, and Cline~\cite{cline_plan_act} -- and the planning mode of the spreadsheet agent Shortcut AI~\cite{shortcut_ai}. We surveyed these tools in June 2026; as they are updated frequently, some of these features may have since changed.

\begin{table*}[t]
\centering
\footnotesize
\renewcommand{\arraystretch}{1.25}
\begin{tabular}{p{0.30\textwidth} *{6}{>{\centering\arraybackslash}p{0.085\textwidth}}}
\toprule
\textbf{Design Feature} & \textbf{Our Prototype} & \textbf{Claude Code} & \textbf{VSCode Agent Mode} & \textbf{Cursor} & \textbf{Cline} & \textbf{Shortcut AI} \\
\midrule
\rowcolor{gray!10}
Read-only access to codebase during planning & \checkmark & \checkmark & \checkmark & \checkmark & \checkmark & \checkmark \\
User-controlled explicit mode switching & \checkmark & \checkmark & \checkmark & \checkmark & \checkmark & \checkmark \\
\rowcolor{gray!10}
Asks clarifying questions before planning & \checkmark & \checkmark & \checkmark & \checkmark & \checkmark & \checkmark \\
\rowcolor{gray!10}
Clarifying-question modality\textsuperscript{a} & NL & MCQ & NL & MCQ & NL & NL \\
Dynamic plan-status updates during execution & \checkmark & \checkmark & \checkmark & \checkmark & \checkmark & \checkmark \\
\rowcolor{gray!10}
Persistent visible plan artifact & \checkmark & \ding{55} & \checkmark & \checkmark & \ding{55} & \ding{55} \\
Direct manipulation of plan steps & \checkmark & \ding{55} & \ding{55} & \ding{55} & \ding{55} & \ding{55} \\
\rowcolor{gray!10}
Partial execution of the plan & \checkmark & \ding{55} & \ding{55} & \ding{55} & \ding{55} & \ding{55} \\
\bottomrule
\end{tabular}
\caption{Relation between the design features of our prototype (Section~\ref{design}) and the Plan Modes of existing tools, as surveyed in June 2026. \checkmark~indicates the tool supports the feature; \ding{55} indicates it does not. 
\textsuperscript{a}Modality of the clarifying-question interaction: NL = natural language in chat; MCQ = multiple-choice questions through UI cards. 
}
\label{tab:design_comparison}
\end{table*}

\subsection{Study Materials}
We include the following study materials: the screener all prospective participants filled out, along with the criteria for qualification for the study; the protocol for the user study; and the descriptions provided to participants about the tasks they worked on.

\subsubsection{Screener Survey}
\label{app:screener}

\begin{enumerate}[label=\arabic*.]
    \item Which of the following apps do you use for professional purposes more than once a week? Select all that apply.
    \begin{itemize}
        \item Notion (\textit{May Select})
        \item Spreadsheet.com (\textit{May Select})
        \item Smartsheet (\textit{May Select})
        \item Adobe Photoshop (\textit{May Select})
        \item Google Docs (\textit{May Select})
        \item Gmail (\textit{May Select})
        \item Microsoft Word (\textit{May Select})
        \item Microsoft Excel (\textit{Qualify})
        \item Google Slides (\textit{May Select})
        \item Microsoft Outlook (\textit{May Select})
        \item Canva (\textit{May Select})
        \item Google Sheets (\textit{May Select})
        \item Microsoft PowerPoint (\textit{May Select})
        \item None on this list (\textit{Disqualify})
    \end{itemize}

    \item How do you most frequently access productivity applications like Word, Excel, or PowerPoint?
    \begin{itemize}
        \item Smartphone app (\textit{Disqualify})
        \item Internet browser (\textit{Disqualify})
        \item Desktop or laptop application (\textit{Qualify})
        \item Tablet (Android/iPad) app (\textit{Disqualify})
    \end{itemize}

    \item How frequently do you use spreadsheets?
    \begin{itemize}
        \item Once a month (\textit{Disqualify})
        \item Once a week (\textit{Qualify})
        \item Multiple times a week (\textit{Qualify})
        \item Never (\textit{Disqualify})
    \end{itemize}

    \item What do you primarily use spreadsheets for? Please select all that apply.
    \begin{itemize}
        \item Viewing information and charts created by others (\textit{May Select})
        \item Creating new spreadsheets with structured/unstructured data (\textit{Qualify})
        \item Editing existing spreadsheets through data entry/visualization (\textit{Qualify})
        \item Performing in-depth data modelling/analysis (\textit{Qualify})
        \item None of the above (\textit{Disqualify})
    \end{itemize}

    \item What is your expertise level with Excel?
    \begin{itemize}
        \item Beginner (Basic data wrangling, sheet formatting) (\textit{Disqualify})
        \item Intermediate (Formulas like SUM, COUNT, MAX, charts, PivotTables) (\textit{Qualify})
        \item Advanced (Power Query, Data Model, formulas like VLOOKUP) (\textit{Qualify})
        \item Expert (Data Model, indirect references, Python formulas, formulas like LET and LAMBDA) (\textit{Qualify})
    \end{itemize}

    \item Have you used/heard of VLOOKUP in Excel before?
    \begin{itemize}
        \item I have used VLOOKUP before (\textit{Qualify})
        \item I have not used VLOOKUP before, but I have heard of it and know what it does (\textit{Qualify})
        \item I have not used VLOOKUP before and I do not know what it does, but I have heard of it (\textit{Disqualify})
        \item I have not used or heard of VLOOKUP before (\textit{Disqualify})
    \end{itemize}

    \item Which of the following would you use to describe yourself? Choose the answer that best fits your scenario.
    \begin{itemize}
        \item A college, graduate, university, or a post-graduate student (\textit{Disqualify})
        \item Employee or Owner of a business with less than 50 employees (\textit{Qualify})
        \item Employee or Owners of a business with 50 or more employees (\textit{Qualify})
        \item Freelance Worker, Gig Worker or Volunteer Worker (\textit{Qualify})
        \item None of the above (\textit{Disqualify})
    \end{itemize}

    \item What is your experience with using AI?
    \begin{itemize}
        \item Extensive -- I use AI-powered tools frequently for work and/or personal tasks (\textit{Qualify})
        \item Occasional -- I have used AI tools but only in limited scenarios (\textit{Qualify})
        \item Rare -- I have only tried AI tools a few times (\textit{Qualify})
        \item None -- I have never used AI for either personal or work tasks (\textit{Disqualify})
    \end{itemize}
\end{enumerate}

\subsubsection{Study Protocol}
\label{app:protocol}
Sessions followed one of two orderings, \textit{Plan Mode First} or \textit{Act Mode First}, counterbalanced across participants. Both followed the same structure: an introduction, a warm-up/tutorial task with the first tool, two 15-minute study tasks (one per tool), and a post-study questionnaire. Here, we provide the full protocol for the ordering where participants used Plan Mode first.

\textbf{Introduction} 

Hi, good morning/afternoon/evening, how are you doing? Thanks for taking time out to participate in this user study.

Today, you will be using two versions of an AI tool in Excel to work on two different spreadsheet tasks by remotely controlling my screen. For each task, try to make as much progress as possible within 15 minutes, after which we will move on to the next task. Before we start, I'll give a short tutorial on how to use the AI tool.

The session will be around 45 minutes long, and you will fill out a post-session survey at the end.

Do you have any questions?

Great, let me share my screen/set up quick assist. (Make sure to share screen either way!)

\textbf{Warm-up Task} 

Let's go through a quick demo task so you are familiar with Plan Mode, the first tool you'll be using today.

This AI tool is a chatbot, similar to something like ChatGPT, except that it has the capability to make complex edits directly to your Excel workbook. Plan mode allows you and the AI to create a plan without making any changes to the workbook before it actually makes any changes. For example, let us ask to create a spreadsheet for reporting grades of students in a class.

\textit{``I want to create a gradebook for my class of grade 10 math students''}

The agent first asks some questions, let's say these are our preferences.

\textit{``numerical grading, 30 students, only 2 exams, rest up to you''}

Based on this, it will build the plan. Now, you can see the plan here. You can directly edit the plan, or revise the plan with the AI through the chat. Once you are satisfied, you can execute all the steps or execute up to any step you want. Pressing execute will leave plan mode and go to act mode, where the AI can make edits to the spreadsheet. You can see the internal reasoning of the process followed by the agent here. You can of course continue the conversation to make any edits.

You can switch back and forth from plan mode whenever you wish, but please try to use plan mode at the beginning, it will be on by default.

Do you have any questions?

For the first task, \textsc{Task Description 1} (Section~\ref{app:task_descriptions}).

Also, please try to think aloud as you work on the task. You can go ahead and start!

\textbf{Participant completes Task 1 (15 minutes)} 

Great! Now, let's move on the second part of the study. For this part, you will be using the Act Mode. In this mode, the AI can start making changes immediately, and will not try to create a plan. So, for example, if we say

\textit{``Create a workbook to report the sales of my furniture company''}

The AI will start to do so immediately.

Do you have any questions?

So for this task, \textsc{Task Description 2} (Section~\ref{app:task_descriptions}).

Again, please try to think aloud as you work through the task.

\textbf{Participant completes Task 2 (15 minutes)} 

\textbf{Post-Task} 

Thanks for participating! The session was very insightful. Now I'm going to open the post-study questionnaire on my screen, please go ahead and fill that out. Make sure to answer the questions based on your experiences with the tool in general, not about the specific task you worked on with the tool, and please talk me through your answers.

\textit{Participant fills the post-study questionnaire.}

Great! Before we end, do you have any questions?

Okay, then once again, thanks for participating! Have a nice day!

\subsubsection{Task Descriptions}
\label{app:task_descriptions}

\begin{itemize}
    \item \textbf{Creation Tasks}
    \begin{itemize}
    \item \textit{Personal Budget:} Try and create your own personal budget, based on your actual needs and preferences. Your objective is to create a useful spreadsheet that you could use practically to track your finances. We will send you the spreadsheet after the session for you to use if you want to.

    \item \textit{Personal Schedule:} Create a personal schedule for yourself; your objective being to manage your time more efficiently. Again, we will send you the spreadsheet after the session for you to use.
    \end{itemize}

    \item \textbf{Analysis Tasks}
    \begin{itemize}
    \item \textit{Movie:} I'm providing you with a movie dataset, and your task is to find the next movie for you to watch based on your own preferences. Let me provide you with a brief summary of the dataset -- it contains information about the genres, languages, runtimes and release dates of the movies along with keywords, taglines, and an overview of the movie. It also has information about the production company, country of origin, budget, revenue, and lifetime popularity of the movie. It also has some audience ratings, with an average rating and number of ratings.

\item \textit{Vacation:} I'm providing you with a dataset of vacation destinations and hotels, and your objective is to plan your next vacation. Again, I'll give a quick summary of the dataset -- it contains information about the specific hotel, such as the type of accommodation, the stars, the rating, number of ratings, distance from city center, and price. It also contains information about the place itself, such as the location, a short description of the attractions, the average monthly temperature, general budget level, and ratings out of 5 for different aspects of the city such as culture, adventure, nightlife, beaches, etc.  
\end{itemize}
\end{itemize}

\subsection{Feature Codebook}
Table \ref{tab:feature_codebook} describes the codebook we used to code features across the Creation tasks in our study.

\begin{table}[t]
\centering
\footnotesize

\begin{tabular}{p{0.32\columnwidth} p{0.60\columnwidth}}
\toprule
\multicolumn{2}{c}{\textbf{Budget Task Codebook}} \\
\midrule
\multicolumn{2}{c}{\textbf{Primary Codes}} \\
\midrule
\textbf{Code} & \textbf{Description} \\
\midrule

\rowcolor{gray!10}
{incomes} & Artifacts dedicated to describing income from any source \\

{expenses} & Artifacts dedicated to describing expenses of any kind \\

\rowcolor{gray!10}
{savings} & Artifacts dedicated to summarizing savings or comparing incomes and expenses \\

\midrule
\multicolumn{2}{c}{\textbf{Secondary Codes}} \\
\midrule
\textbf{Code} & \textbf{Description} \\
\midrule

\rowcolor{gray!10}
{expected vs actual} & Comparing expected (budgeted) amounts vs.\ actual amounts \\

{individual} & Noting down individual expenses [only for expenses code] \\

\rowcolor{gray!10}
{financial goals} & Describing savings goals and/or progress [only for savings code] \\

{time-wise comparison} & Comparing amounts across any quantity of time (e.g., monthly, weekly, yearly, daily) \\

\rowcolor{gray!10}
{categories} & Comparing amounts across categories \\

{specific category} & Deep-dive into amounts for one specific category \\

\rowcolor{gray!10}
{taxes} & Artifacts dedicated to tax calculations \\

{balances} & Artifacts dedicated to account balance calculations [only for savings code] \\

\rowcolor{gray!10}
{tips} & AI-generated tips for budget management/interpretations of spreadsheet data \\

\bottomrule
\end{tabular}

\vspace{1em}

\begin{tabular}{p{0.32\columnwidth} p{0.60\columnwidth}}
\toprule
\multicolumn{2}{c}{\textbf{Schedule Task Codebook}} \\
\midrule
\multicolumn{2}{c}{\textbf{Primary Codes}} \\
\midrule
\textbf{Code} & \textbf{Description} \\
\midrule

\rowcolor{gray!10}
{calendar} & Calendar-like table describing tasks \\

{metainfo} & Information about the workbook itself such as legends \\

\rowcolor{gray!10}
{summary} & Analysis of task distribution over time \\

{task list} & Table tracking all tasks, not formatted as a calendar \\

\rowcolor{gray!10}
{tips} & AI-generated tips for time management \\

\midrule
\multicolumn{2}{c}{\textbf{Secondary Codes}} \\
\midrule
\textbf{Code} & \textbf{Description} \\
\midrule

\rowcolor{gray!10}
{priority} & Contains information about task priorities \\

{status} & Contains information about task statuses (pending, in progress, etc.) \\

\rowcolor{gray!10}
{color coding} & Color coding indicating task type/status [only for calendar code] \\

{specific} & Tracking individual, non-repeating tasks \\

\rowcolor{gray!10}
{task categories} & Contains information about the categories of different tasks \\

{instructions} & Instructions on how to use the spreadsheet [only for metainfo code] \\

\bottomrule
\end{tabular}

\caption{Hierarchical feature codebooks used in the analysis.}
\label{tab:feature_codebook}
\end{table}

\subsection{Modifications to Post-Study Questionnaire}
Questions related to `Immersiveness' in the original scale (e.g., \textit{``My attention was fully tuned to the activity, and I forgot about the system or tool that I was using.''}) were less applicable to agentic tools that make complex edits than to purely co-creative tools. We therefore adapted these items to more directly measure participant attention across the tool and the task (e.g., ``My attention was more focused on the tool'' and ``My attention was more focused on the task'').

We also modified the phrasing of the `Contribution' statement (\textit{``I/ the system made the sketch''}) to capture creative responsibility (\textit{``I/ the agent is creatively responsible for the spreadsheet''}) rather than literal contribution since the agent performed all of the spreadsheet actions in the task. To measure whether working with the agent helped participants push beyond their capabilities to produce tables they could not create by themselves, we added an extra question to capture participant perceptions on their \textit{Capability} with respect to the final output; this dimension was not originally captured by the questionnaire as it focuses on purely creative tasks. These edits are not meant to support new statistical claims, but rather to contextualise our qualitative findings on participants' interaction patterns across modes.

\subsection{Additional Figures} \label{app:additional_figures}
In this section, we include some figures and tables we did not have space for in the main draft.

\begin{figure}[t]
    \centering

    \begin{subfigure}[t]{\linewidth}
        \centering
        \includegraphics[width=\linewidth]{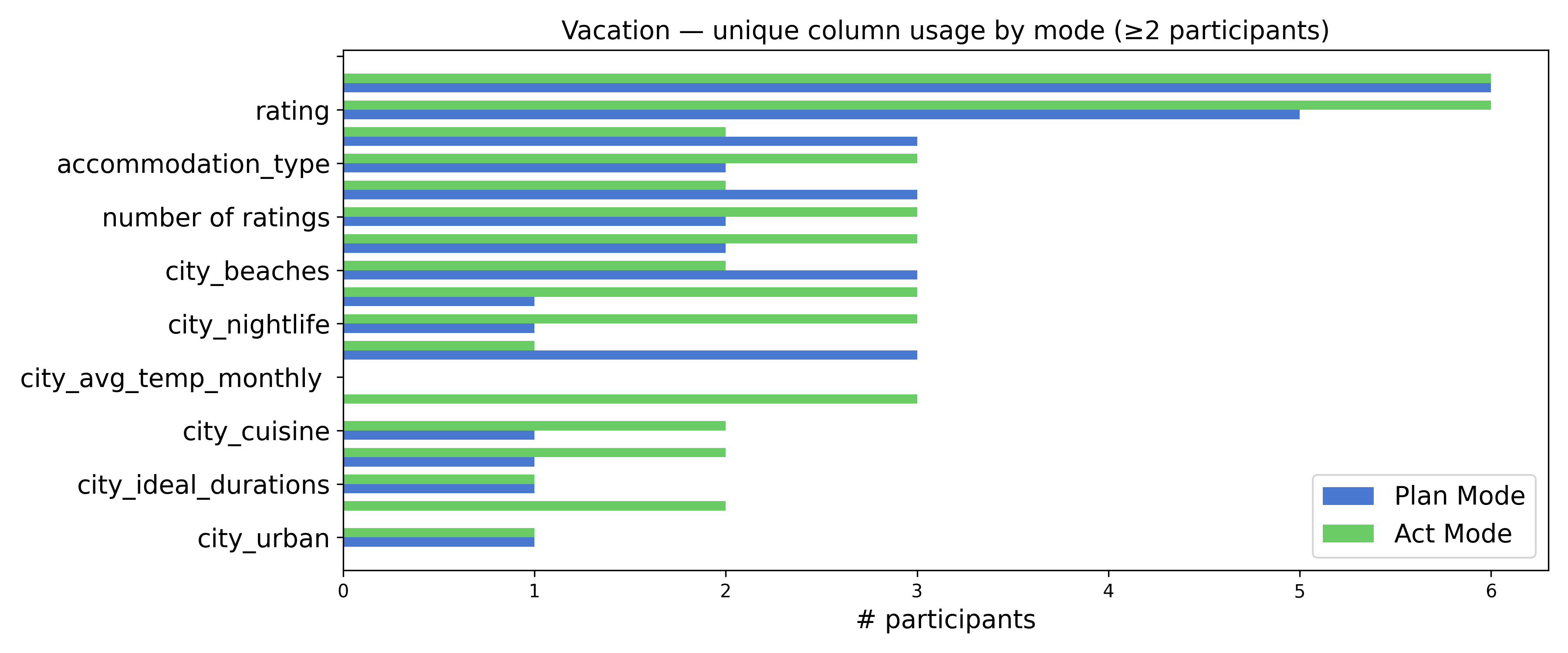}
        \caption{Columns used by participants in the Vacation task}
        \label{fig:sub11}
    \end{subfigure}

    \vspace{0.5em}

    \begin{subfigure}[t]{\linewidth}
        \centering
        \includegraphics[width=\linewidth]{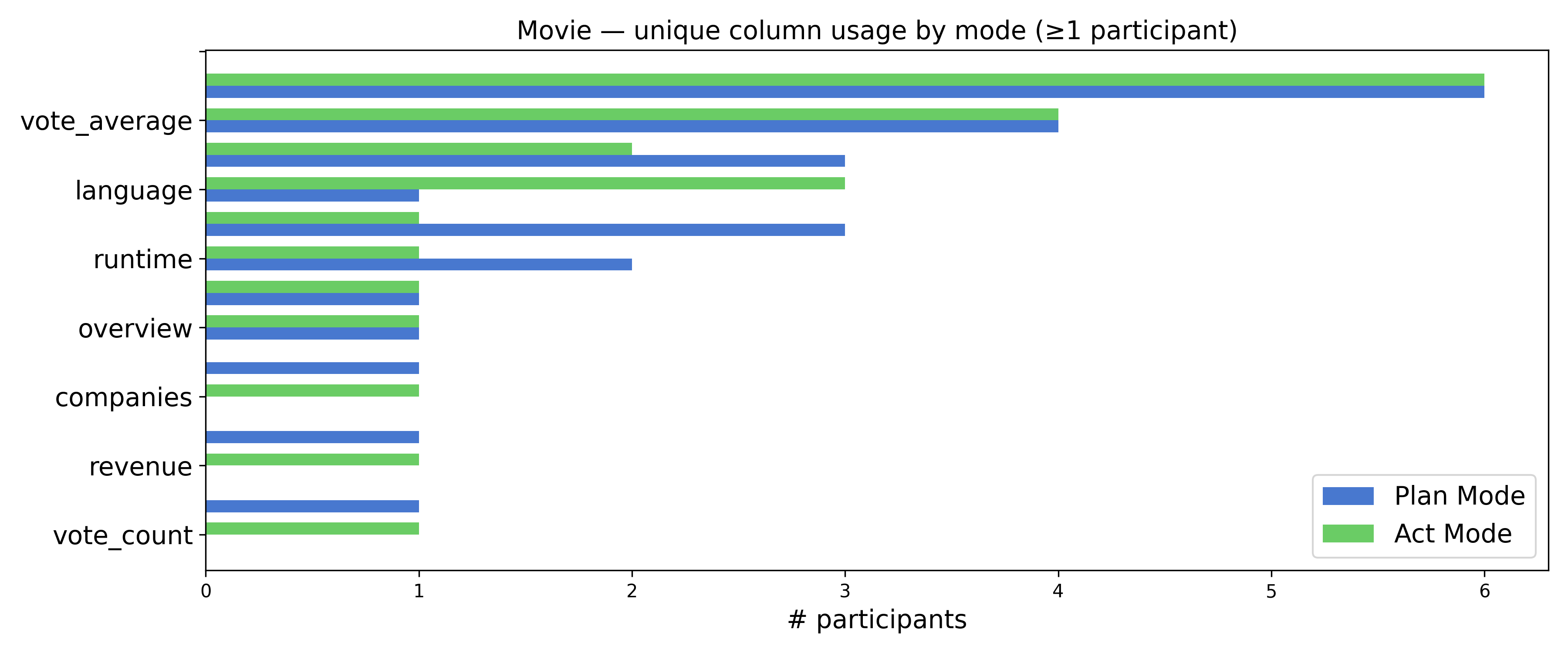}
        \caption{Columns used by participants in the Movie task}
        \label{fig:sub22}
    \end{subfigure}

    \caption{Columns used by participants in the Analysis tasks across modes}
    \label{fig:column_usage}
\end{figure}

\begin{figure}[t]
    \centering
    \includegraphics[width=\linewidth]{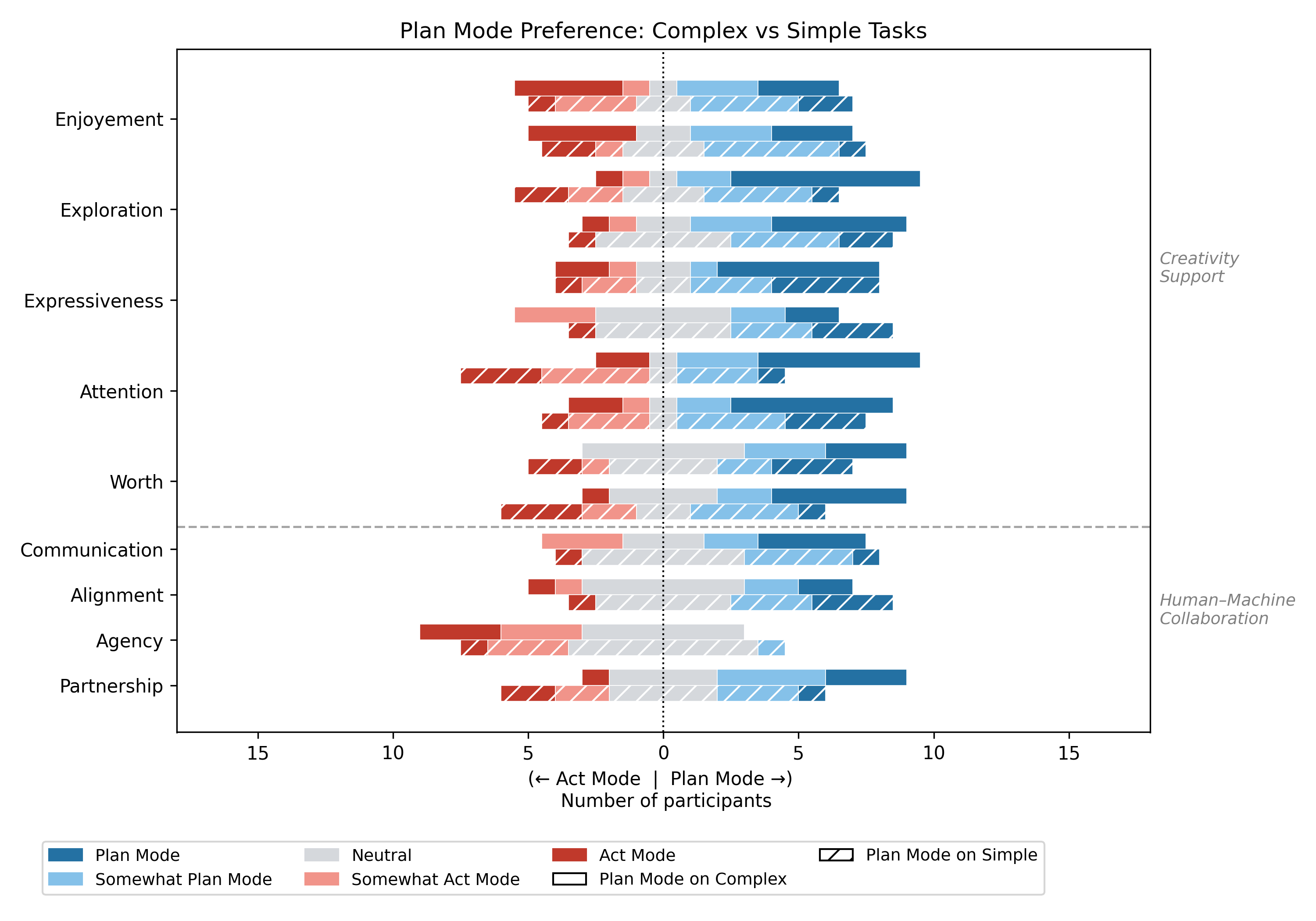}
    \caption{Participant preferences across task complexity.}
    \label{fig:questionnaire_complexity}
\end{figure}

\begin{figure}[t]
    \centering
    \includegraphics[width=\linewidth]{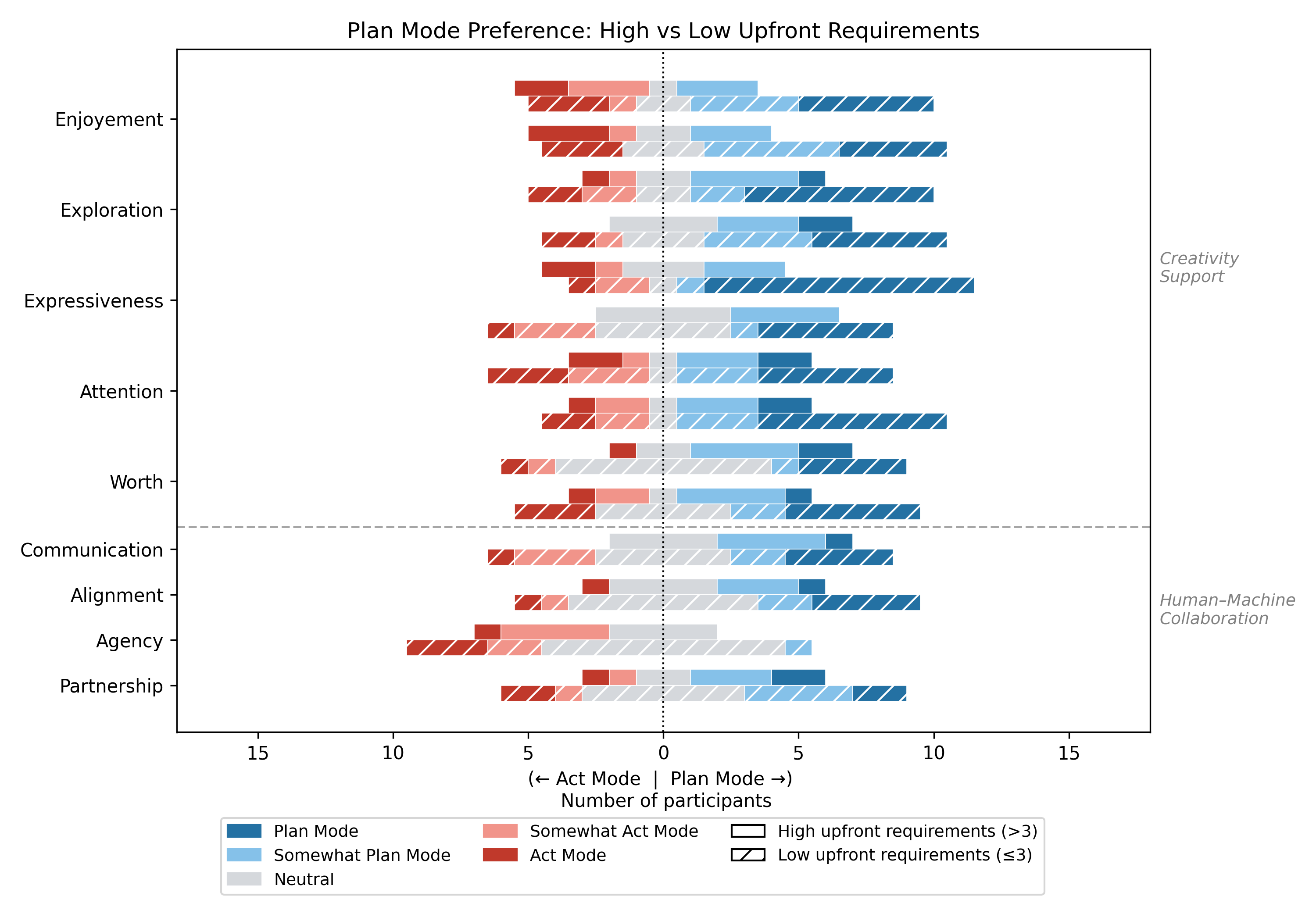}
    \caption{Participant preferences based on upfront requirements.}
    \label{fig:plan_mode_preference_by_upfront}
\end{figure}

Figure \ref{fig:column_usage} shows, for each of the two \Analysis tasks, how many participants referenced each information column of the provided dataset while working in Plan Mode versus Act Mode, and Table \ref{tab:filter_sort_usage} shows the average number of these columns participants used for filtering, sorting, and both across modes and tasks.
Figure \ref{fig:questionnaire_complexity} and Figure \ref{fig:plan_mode_preference_by_upfront} show participants' post-study questionnaire preferences for Plan Mode versus Act Mode compared across the complexity of the tasks they worked on and the upfront requirements they provided for their tasks, respectively. 

Table \ref{tab:mode_tasktype} shows the mean number of requirements participants expressed per task, broken down by mode (Plan Mode vs.\ Act Mode) and task type (\Analysis vs.\ \Creation), while Table \ref{tab:mode_task} provides the same breakdown at the level of each of the four individual tasks (\Budget, \Schedule, \Vacation, \Movie).

\begin{table}[t]
\centering
\begin{tabular}{lcc}
\toprule
 & Analysis & Create \\
\midrule
Plan Mode & 8.6 & 9.8 \\
Act Mode  & 7.4 & 7.8 \\
\bottomrule
\end{tabular}
\caption{Mean number of requirements by mode and task type.}
\label{tab:mode_tasktype}
\end{table}

\begin{table}[t]
\centering
\begin{tabular}{lcccc}
\toprule
 & Budget & Schedule & Vacation & Movie \\
\midrule
Plan Mode & 10.7 & 9.0 & 10.2 & 7.0 \\
Act Mode  & 9.3  & 6.2 & 9.2  & 5.7 \\
\bottomrule
\end{tabular}
\caption{Mean number of requirements by mode and task.}
\label{tab:mode_task}
\end{table}

Table \ref{tab:followup_reason} shows the origin of participants' follow-up requirements across modes.

\begin{table}[t]
\centering
\begin{tabular}{lcccc}
\toprule
 & Proactive & Clarification & Plan & Fix \\
\midrule
Plan Mode & 6\% (3) & 82\% (40) & 6\% (3) & 6\% (3) \\
Act Mode  & 12\% (2) & 12\% (2) & 0\% (0) & 75\% (12) \\
\bottomrule
\end{tabular}
\caption{Distribution of follow-up requirements by origin within each mode. Percentages are shown with counts in parentheses.}
\label{tab:followup_reason}
\end{table}

\begin{table}[t]
\centering

\begin{subtable}{\linewidth}
\centering
\begin{tabular}{lccc}
\toprule
 & Both Tasks & Vacation & Movie \\
\midrule
Plan Mode & 4.3 & 5.2 & 3.3 \\
Act Mode  & 4.1 & 5.3 & 2.8 \\
\bottomrule
\end{tabular}
\caption{Filtering}
\label{tab:filters}
\end{subtable}

\vspace{0.75em}

\begin{subtable}{\linewidth}
\centering
\begin{tabular}{lccc}
\toprule
 & Both Tasks & Vacation & Movie \\
\midrule
Plan Mode & 2.3 & 3.5 & 1.0 \\
Act Mode  & 2.6 & 3.8 & 1.3 \\
\bottomrule
\end{tabular}
\caption{Sorting}
\label{tab:sorts}
\end{subtable}

\begin{subtable}{\linewidth}
\centering
\begin{tabular}{lccc}
\toprule
 & Both Tasks & Vacation & Movie \\
\midrule
Plan Mode & 5.3 & 6.5 & 4.0 \\
Act Mode  & 5.8 & 8.0 & 3.7 \\
\bottomrule
\end{tabular}
\caption{Filtering/Sorting}
\label{tab:sorts_and_filter}
\end{subtable}

\caption{Average number of unique information columns used per analysis task across modes}
\label{tab:filter_sort_usage}
\end{table}

\end{document}